\documentclass[%
reprint,
superscriptaddress,
amsmath, amssymb,
aps,
prb,
]{revtex4-2}

\usepackage[cp1252, utf8]{inputenc}
\usepackage{standalone}
\usepackage{graphicx}
\usepackage{bm}
\usepackage{braket}
\usepackage{dsfont}
\usepackage[%
colorlinks=true,
urlcolor=blue,
linkcolor=blue,
citecolor=blue
]{hyperref}
\usepackage{orcidlink}

\graphicspath{{images/}}

\newcommand\imglabelsize{16} 

\begin{document}

\title{Computing Floquet quasienergies in finite and extended systems: Role of electromagnetic and quantum-geometric gauges}

\author{{\'A}lvaro R. Puente-Uriona\,\orcidlink{0000-0003-1915-8804}}
\email[]{alvaro.ruiz@ehu.eus}
\affiliation{Centro de F{\'i}sica de Materiales, Universidad del Pa{\'i}s Vasco (UPV/EHU), 20018 San Sebasti{\'a}n, Spain}

\author{Michele Modugno\,\orcidlink{0000-0002-0532-1423}}
\affiliation{Department of Physics, University of the Basque Country UPV/EHU, 48080 Bilbao, Spain}
\affiliation{IKERBASQUE, Basque Foundation for Science, 48009 Bilbao, Spain}
\affiliation{EHU Quantum Center, University of the Basque Country UPV/EHU, 48940 Leioa, Biscay, Spain}

\author{Ivo Souza\,\orcidlink{0000-0001-9901-5058}}
\affiliation{Centro de F{\'i}sica de Materiales, Universidad del Pa{\'i}s Vasco (UPV/EHU), 20018 San Sebasti{\'a}n, Spain}
\affiliation{IKERBASQUE, Basque Foundation for Science, 48009 Bilbao, Spain}

\author{Julen Iba\~{n}ez-Azpiroz\,\orcidlink{0000-0003-0922-0926}}
\affiliation{Centro de F{\'i}sica de Materiales, Universidad del Pa{\'i}s Vasco (UPV/EHU), 20018 San Sebasti{\'a}n, Spain}
\affiliation{IKERBASQUE, Basque Foundation for Science, 48009 Bilbao, Spain}
\affiliation{Donostia International Physics Center (DIPC), 20018 Donostia-San Sebasti\'an, Spain}

\date{\today}

\begin{abstract}

We present an approach to compute the Floquet quasienergy spectrum of time periodic systems. The method allows to characterize the light-matter interaction in finite and extended structures by carefully addressing the resolution of the position operator. In periodic systems we discuss the role of the quantum-geometric gauge freedom of Bloch states and employ a Wannier based scheme to compute the required matrix elements. As a consequence, the method is accurate and applicable to a broad range of systems, from atoms and molecules to cold atomic gases and materials described by density functional theory, as well as model systems. We demonstrate the applicability of the approach by studying two cases: a particle trapped in a one dimensional box and the semiconducting material BC$_2$N. We employ the first example to provide a numerical proof of the invariance of the Floquet quasienergy spectrum with respect to the choice of electromagnetic gauge. The analysis of BC$_2$N then serves to illustrate the physical effects described by the quasienergies, such as multiphoton resonances, and their expected range of occurrence in real materials in terms of external electric field and frequency of the drive pulse.

\end{abstract}

\maketitle

\section{Introduction}

The Floquet theory~\cite{sambeSteadyStatesQuasienergies1973, chuGeneralizedFloquetTheoretical2007} offers a versatile theoretical tool for studying the time evolution of periodically driven systems. With its aid, optical phenomena like multiphoton resonances~\cite{fainshteinGeneralPropertiesQuasienergetic1978, weinbergMultiphotonInterbandExcitations2015} or dynamic localization~\cite{dunlapDynamicLocalizationCharged1986, keshavamurthyDynamicalTunnelingTheory2011, arlinghausDynamicLocalizationOptical2011, eckardtColloquiumAtomicQuantum2017} can be naturally described. Additionally, the Floquet formalism is also useful in the characterization of nontrivial phases~\cite{kitagawaTopologicalCharacterizationPeriodically2010, nakagawaWannierRepresentationFloquet2020, puente-urionaTopologicalPhaseDiagram2024}, such as the fractional Chern insulator~\cite{grushinFloquetFractionalChern2014}, or the geometric interpretation of nonlinear optical responses~\cite{morimotoTopologicalNatureNonlinear2016}.

Over the past two decades, the Floquet theory has been intensively employed to elucidate the behavior of cold atoms, where periodically modulated optical lattices simulate the effects of time dependent forces~\cite{goldmanPeriodicallyDrivenQuantum2014, eckardtColloquiumAtomicQuantum2017,bukovUniversalHighfrequencyBehavior2015, castroFloquetEngineeringQuantum2023}. In this field, Floquet theory proves instrumental in understanding the creation of synthetic electromagnetic fields~\cite{jotzuExperimentalRealizationTopological2014, modugnoCorrespondenceShakenHoneycomb2017}, the interaction-driven transition from bosonic superfluids to Mott insulator phases~\cite{eckardtSuperfluidInsulatorTransitionPeriodically2005, zenesiniCoherentControlDressed2009} or the realization of dynamical structures such as Majorana fermions~\cite{jiangMajoranaFermionsEquilibrium2011}, among other effects.

Over the past decade, the use of the Floquet theory has steadily increased in the field of condensed matter. The experimental discovery of Floquet-Bloch states~\cite{wangObservationFloquetBlochStates2013} has paved the way for the engineering of topological materials whose properties can be controlled by periodic pulses~\cite{hubenerCreatingStableFloquet2017, itoBuildupDephasingFloquet2023, delatorreColloquiumNonthermalPathways2021}. The advent of intense ultrafast laser pulse techniques~\cite{beaurepaireUltrafastSpinDynamics1996, koopmansUltrafastMagnetoOpticsNickel2000} has furthermore allowed the manipulation of properties of solids with few-cycle fields, including the spin, orbital and vibrational degrees of freedom~\cite{eschenlohrUltrafastSpinTransport2013, waldeckerMomentumResolvedViewElectronPhonon2017, shallcrossUltrafastOpticalControl2022, sharmaValleyControlLinearly2022, schulerLocalBerryCurvature2020}.

Despite progress on multiple theoretical aspects of the Floquet formalism, the growing complexity of the target systems, from cold atoms to condensed matter, requires the development and deployment of improved numerical tools as well. A parameter-free approach, capable of accounting for real-world systems under complex driving forces is necessary for an accurate description of the dynamics accessed in experiments. A few recent works have taken steps in this direction by incorporating concepts of the Floquet formalism into first principles schemes commonly employed in density functional theory (DFT)~\cite{hubenerCreatingStableFloquet2017, alliatiFloquetFormulationDynamical2023}.

In this manuscript we provide a general numerical approach to compute two central objects in the Floquet formalism, namely the \emph{effective Floquet Hamiltonian} and its eigenvalues, the so-called \emph{quasienergies}. These describe relevant time dependent phenomena such as the aforementioned multiphoton resonances and dynamical localization, and serve as a basis for multiple treatments within the Floquet theory~\cite{bukovUniversalHighfrequencyBehavior2015, holthausFloquetEngineeringQuasienergy2016}. To this end, we develop a scheme based on Wannier functions that handles arbitrary driving forces and is applicable to a broad range of systems, from simple theoretical models to real-world materials described by DFT. This tool is therefore applicable to systems that are commonly studied in the cold atom and condensed matter communities.

The paper is organized as follows. First in Sec.~\ref{sec:em-gauge} we review the basic properties of the Floquet formalism and formally prove the electromagnetic gauge-covariance (invariance) of the effective Floquet Hamiltonian (quasienergies). We provide a numerical verification in Sec.~\ref{sec:particle_in_a_box} by considering the length and velocity gauges in the benchmark particle in a box system. In Sec.~\ref{sec:extended_systems} we generalize the formalism to deal with extended systems, where we pay attention to the intricacies related to the position operator and the quantum-geometric gauge freedom~\cite{blountFormalismsBandTheory1962}. We exemplify the use of our approach in the semiconducting material BC$_2$N in Sec.~\ref{sec:real_materials}, and discuss relevant features of the quasienergy spectrum, such as avoided crossings between energy levels. The most technical parts of the work are described in the appendices.

\section{Time evolution operator, Floquet Hamiltonian and Electromagnetic Gauge transformations}\label{sec:em-gauge}

\subsection{Preliminaries}\label{sec:prelim}

Consider the Hamiltonian operator in absence of external perturbations:
\begin{equation}\label{eq:basic_hamil}
\hat{H}_0 = \frac{\hat{\bm{p}}^2}{2m} + \hat{V}\left(\hat{\bm{r}}\right).
\end{equation}
If we now apply a time dependent electromagnetic field, the system is described by the minimal coupling Hamiltonian~\cite{cohen-tannoudjiQuantumMechanics1977}
\begin{equation}\label{eq:formal_hamil}
\hat{H}(t) = \frac{\left[\hat{\bm{p}} - q\hat{\bm{A}}(\hat{\bm{r}}, t)\right]^2}{2m} + \hat{V}\left(\hat{\bm{r}}\right) + q\hat{\phi}(\hat{\bm{r}}, t),
\end{equation}
where the operators $\hat{\bm{A}}(\hat{\bm{r}}, t)$ and $\hat{\phi}(\hat{\bm{r}}, t)$ are the vector and scalar potentials, respectively. The Schr{\"o}dinger equation for this Hamiltonian reads
\begin{equation}
i\hbar\frac{\partial}{\partial t}\ket{\psi(t)} = \hat{H}(t)\ket{\psi(t)}.
\end{equation}

The dynamic evolution of the wave function from an initial time $t_{\text{s}}$ to a final time $t$ can be expressed making use of the time evolution operator $\hat{U}(t,t_{\text{s}})$:
\begin{equation}\label{eq:def_teo}
\ket{\psi(t)} = \hat{U}(t, t_{\text{s}})\ket{\psi(t_{\text{s}})}.
\end{equation}
This quantity can be formally defined using the time dependent Hamiltonian in Eq.~\eqref{eq:formal_hamil} as~\cite{sakuraiModernQuantumMechanics2017}
\begin{equation}\label{eq:def_tevop}
\hat{U}(t, t_{\text{s}}) = \text{exp}\left[\frac{-i}{\hbar}\int_{t_{\text{s}}}^{t}\hat{H}(\tau)d\tau\right],
\end{equation}
and it obeys the semigroup property
\begin{equation}\label{eq:semigroup}
\hat{U}(t, t_{\text{s}}) = \hat{U}(t, t')\hat{U}(t', t_{\text{s}}),\quad\forall t'\in(-\infty, \infty).
\end{equation}

\subsection{The Floquet picture}\label{sec:floq}

Let us now consider the time dependent Hamiltonian of Eq.~\eqref{eq:formal_hamil} satisfying a time periodic constraint of the form
\begin{equation}\label{eq:floquet_condition}
\hat{H}(t) = \hat{H}(t + T),\;T = 2\pi/\omega,
\end{equation}
where $T$ defines the period and $\omega$ its associated frequency. Any instant in time can be decomposed as $t = nT + t_r, \; n \in \mathbb{Z}$ with $t_r\in[t_{\text{s}}, t_{\text{s}}+T]$. Then, making use of the semigroup property in Eq.~\eqref{eq:semigroup}, one can rewrite the time evolution operator as
\begin{equation}\label{eq:per_tev}
\hat{U}(t, t_{\text{s}}) = \hat{U}(t_r, t_{\text{s}})\left[\hat{U}(t_{\text{s}} + T, t_{\text{s}}) \right]^n.
\end{equation}
The quantity inside brackets $\hat{U}(t_{\text{s}} + T, t_{\text{s}})$ is the \emph{one-cycle evolution operator}; just as the translation operator by a lattice vector plays a prominent role in condensed-matter physics, so does $\hat{U}(t_{\text{s}} + T, t_{\text{s}})$ in the Floquet formalism~\cite{holthausFloquetEngineeringQuasienergy2016}. Since it is a unitary operator, one can define the following Hermitian operator
\begin{equation}\label{eq:eff_floq_hamil}
\hat{H}_{\text{F}}(t_{\text{s}}) = i\frac{\hbar}{T} \text{log}\left[ \hat{U}(t_{\text{s}} + T, t_{\text{s}}) \right]\end{equation}
known as the \emph{effective Floquet Hamiltonian}. Its eigenvalues $\varepsilon_{n}$ are real and are commonly termed \emph{Floquet quasienergies}. Because of the presence of the complex logarithm, these eigenvalues are not uniquely defined; if $\varepsilon_{n}$ is a valid eigenvalue of $\hat{H}_{\text{F}}(t_{\text{s}})$, then the family of eigenvalues
\begin{equation}\label{eq:quasienergy}
\varepsilon_{n,\{m\}} = \varepsilon_{n} + m\hbar\omega, \quad m \in\mathbb{Z},
\end{equation}
is equally valid; hence, the integer $m$ defines an entire class of equivalent representatives~\cite{holthausFloquetEngineeringQuasienergy2016}. In view of this, one can define a \emph{quasienergy first Brillouin zone}~\cite{sambeSteadyStatesQuasienergies1973} as
\begin{equation}\label{eq:quasienergy_BZ}
\varepsilon_{n}\in\frac{\hbar\omega}{2}\left[-1, 1\right],
\end{equation}
to fix the difference between the eigenvalues of $\hat{H}_{\text{F}}$, that is, the arbitrary integers $m$ in Eq.~\eqref{eq:quasienergy}.

\subsection{Electromagnetic gauge transformations}

\subsubsection{Review of the general case}

Next, we move on to consider the effect of electromagnetic gauge transformations on the central Floquet quantities. Let us first briefly review the general case; the electric and magnetic fields are given by
\begin{equation}
\label{eq:eb_field}
\hat{\bm{E}}(t) = -\bm{\nabla}\hat{\phi}(t) - \frac{\partial \hat{\bm{A}} (t)}{\partial t}, \;
\hat{\bm{B}}(t) = \bm{\nabla}\times \hat{\bm{A}}(t),
\end{equation}
in terms of the vector and scalar potentials. Under a gauge transformation, these potentials transform as
\begin{equation}
\label{eq:gauge_trn_A}
\hat{\bm{A}'}(t) = \hat{\bm{A}}(t) + \bm{\nabla}\hat{\Lambda}(t),\;
\hat{\phi}'(t) = \hat{\phi}(t) - \frac{\partial \hat{\Lambda}(t)}{\partial t},
\end{equation}
with $\hat{\Lambda}(t)\equiv\hat{\Lambda}(\hat{\bm{r}}, t)$ an arbitrary differentiable function. To lighten notation, we are dropping the dependence of $\hat{\Lambda}$, the potentials, and the electric and magnetic fields on $\hat{\bm{r}}$. As required, the above transformations leave the electric and magnetic fields invariant. The state vectors corresponding to two different gauges are related by the unitary transformation
\begin{equation}\label{eq:gauge_trn}
\ket{\psi^{\prime}(t)} = \text{exp}\left[\frac{iq}{\hbar}\hat{\Lambda}(t)\right]\ket{\psi(t)}.
\end{equation}
Then, the invariance of the Schr{\"o}dinger equation implies that the Hamiltonian transforms to the primed gauge as~\cite{cohen-tannoudjiQuantumMechanics1977}
\begin{equation}\label{eq:hamil_trans_prop}
\hat{H'} = \text{exp}\left[\frac{iq}{\hbar}\hat{\Lambda}( t)\right]\hat{H}\;\text{exp}\left[\frac{-iq}{\hbar}\hat{\Lambda}( t)\right] -q \frac{d\hat{\Lambda}(t)}{dt}.
\end{equation}
Considering the relation of Eq.~\eqref{eq:gauge_trn} between the state vectors in different gauges, the time evolution operator in Eq.~\eqref{eq:def_teo} is seen to transform as
\begin{equation}\label{eq:tevop_trans_prop}
\hat{U'}(t, t_{\text{s}}) = \text{exp}\left[\frac{iq}{\hbar}\hat{\Lambda}( t)\right]\hat{U}(t, t_{\text{s}})\;\text{exp}\left[\frac{-iq}{\hbar}\hat{\Lambda}( t_{\text{s}})\right].
\end{equation}
An additional proof of the above equation employing a factorization theorem was provided by Kobe and Yang~\cite{kobeGaugeTransformationTimeevolution1985}.

\subsubsection{Gauge-covariance}\label{sec:gauge-covariance}

In quantum mechanics, an operator $\hat{O}$ transforms covariantly if it transforms as
\begin{equation}\label{eq:generic-O}
\hat{O}^{\prime} = \text{exp}\left[\frac{iq}{\hbar}\hat{\Lambda}( t)\right]\hat{O}\;\text{exp}\left[\frac{-iq}{\hbar}\hat{\Lambda}( t)\right]
\end{equation}
under a gauge transformation. The eigenvalues of gauge-covariant operators are gauge-invariant~\cite{cohen-tannoudjiQuantumMechanics1977}.

As an example, the mechanical momentum $\hat{\bm{\Pi}}=\hat{\bm{p}}-e\hat{\bm{A}}$ follows the transformation relation of Eq.~\eqref{eq:generic-O}, while the canonical momentum $\bm{\hat{p}}$ does not. In general, the Hamiltonian is \emph{not} a gauge-covariant operator due to the presence of the last term in Eq.~\eqref{eq:hamil_trans_prop}; only in the time independent case does it transform covariantly, owing to $d\hat{\Lambda}/{dt}=0$, and the time independent energy eigenvalues then are gauge-invariant. As for the time evolution operator, in general it is not covariant due to the different time factors entering the exponentials in Eq.~\eqref{eq:tevop_trans_prop}. However, this equation does ensure that its matrix elements are gauge-invariant, i.e., the probablity amplitude $\braket{\psi(t)|\hat{U}(t,t_{\text{s}})|\psi(t_{\text{s}})}= \braket{\psi^{\prime}(t)|\hat{U}^{\prime}(t,t_{\text{s}})|\psi^{\prime}(t_{\text{s}})}$ does not depend on the choice of gauge~\cite{kobeGaugeTransformationTimeevolution1985}.

In the following, we show that in time periodic systems the one-cycle evolution operator and, by extension, the Floquet Hamiltonian transform gauge-covariantly, which implies that the Floquet quasienergies are gauge-independent quantities.

\subsubsection{Time periodicity and the Floquet invariant}

Let us reconsider the time periodic condition of Eq.~\eqref{eq:floquet_condition}. For it to hold, both the vector- and scalar-gauge potentials must be time periodic,
\begin{subequations}\label{eq:floq_cond_g_potentials}
\begin{align}
\hat{\bm{A}}(t + T) &= \hat{\bm{A}}(t), \\
\hat{\phi}(t + T) &= \hat{\phi}(t).
\end{align}
\end{subequations}
These conditions, in turn, impose a time periodic restriction on the gauge transformation operator $\hat{\Lambda}(t)$ by virtue of Eq.~\eqref{eq:gauge_trn_A}:
\begin{equation}\label{eq:periodic_lambda_identity}
\hat{\Lambda}(t + T) = \hat{\Lambda}(t).
\end{equation}
Equation~\eqref{eq:periodic_lambda_identity} has some subtle consequences involving the physical conditions for the realization of Floquet compatible driving. We analyze this in Appendix~\ref{app:no_net_momentum_tranfer}. This also has immediate consequences for the gauge transformation property of the one-cycle evolution operator $\hat{U}(t_{\text{s}} + T, t_{\text{s}})$ discussed in Sec.~\ref{sec:floq}. Setting $t=t_{\text{s}}+T$ in Eq.~\eqref{eq:tevop_trans_prop} we obtain
\begin{equation}\label{eq:one_per_tev_trans}
\begin{split}
&\hat{U}'(t_{\text{s}} + T, t_{\text{s}}) \\
& = \text{exp}\left[\frac{iq}{\hbar}\hat{\Lambda}(t_{\text{s}} + T)\right]\hat{U}(t_{\text{s}} + T, t_{\text{s}})\;
\text{exp}\left[\frac{-iq}{\hbar}\hat{\Lambda}(t_{\text{s}})\right] \\
& = \text{exp}\left[\frac{iq}{\hbar}\hat{\Lambda}(t_{\text{s}})\right]\hat{U}(t_{\text{s}} + T, t_{\text{s}})\;
\text{exp}\left[\frac{-iq}{\hbar}\hat{\Lambda}(t_{\text{s}})\right],
\end{split}
\end{equation}
where we used Eq.~\eqref{eq:periodic_lambda_identity} in the last line. The above describes a covariant transformation of the type of Eq.~\eqref{eq:generic-O}. Going one step further and considering the definition of the effective Floquet Hamiltonian in Eq.~\eqref{eq:eff_floq_hamil}, we obtain a similar expression for this related quantity:
\begin{equation}\label{eq:eff_floq_hamil_trans_prop}
\hat{H}'_{\text{F}}(t_{\text{s}}) = \text{exp}\left[\frac{iq}{\hbar}\hat{\Lambda}(t_{\text{s}})\right]\hat{H}_{\text{F}}(t_{\text{s}})\;\text{exp}\left[\frac{-iq}{\hbar}\hat{\Lambda}(t_{\text{s}})\right],
\end{equation}
where we used the property $\text{log}(\hat{W}\hat{Q}\hat{W}^{\dagger}) = \hat{W}\text{log}(\hat{Q})\hat{W}^{\dagger}$ that holds for unitary operators $\hat{Q}$ and $\hat{W}$~\footnote{We first remark that all unitary operators are normal, so the spectral theorem applies. This means that $\hat{Q}$ can be decomposed as $\hat{U}\hat{D}\hat{U}^{\dagger}$, where $\hat{U}$ is unitary, and $\hat{D}$ diagonal. Since $\hat{Q}$ is unitary, it is trivial to show that its eigenvalues, the entries of $\hat{D}$, are complex numbers in the complex unit circle. With this remark in place, we argue that $\hat{W}\hat{Q}\hat{W}^{\dagger}$ is again unitary with eigenvalues in the complex unit circle. This guarantees that the Taylor expansion for the complex logarithm is convergent. Lastly, some algebra shows that
\begin{equation}
\begin{split}
&\text{log}(\hat{W}\hat{Q}\hat{W}^{\dagger}) = \sum_{k=1}^{\infty}\frac{(-1)^{k+1}}{k}\left(\hat{W}\hat{Q}\hat{W}^{\dagger} - \hat{\mathds{1}}\right)^k = \\
& \sum_{k=1}^{\infty}\frac{(-1)^{k+1}}{k}\left(\hat{W}\hat{Q}\hat{W}^{\dagger} - \hat{W}\hat{W}^{\dagger}\right)^k = \\
&\hat{W}\left[\sum_{k=1}^{\infty}\frac{(-1)^{k+1}}{k}\left(\hat{Q} - \hat{\mathds{1}}\right)^k\right]\hat{W}^{\dagger} = \hat{W}\text{log}(\hat{Q})\hat{W}^{\dagger}
\end{split}
\end{equation}
where we have made extensive use of the unitarity property of $\hat{W}$.}. Eq.~\eqref{eq:eff_floq_hamil_trans_prop} shows that the Floquet Hamiltonian is gauge-covariant. This result then establishes that the Floquet quasienergies of Eq.~\eqref{eq:quasienergy} are invariant with respect to the choice of gauge. In the following section, we provide an explicit numerical example of this assertion.

We conclude this section with a brief remark on the gauge-invariance of the quasienergies. The time periodic condition of Eq.~\eqref{eq:periodic_lambda_identity} is not the most general solution to the functional dependence problem of Eq.~\eqref{eq:floq_cond_g_potentials}. The general solution is given by $\hat{\Lambda}(t + T) = \hat{\Lambda}(t) + \hat{\mathds{1}}\beta t$, where $\beta$ is a real constant with units of energy. In practice, $\beta$ acts as a rigid shift of the zero of the gauge potentials, which implies that only \emph{differences} in quasienergies are gauge-invariant under a shift of the zero of energy. With this remark in place, given that $\beta$ is completely irrelevant for the dynamics and that the quasienergy spectrum is only uniquely defined within the quasienergy first Brillouin zone of Eq.~\eqref{eq:quasienergy_BZ}, we will disregard it in the rest of the work and refer to quasienergies as gauge-invariant quantities.

\section{Particle in a box}\label{sec:particle_in_a_box}

Consider the textbook example of a periodically driven particle in a box: a particle with mass $m$ and charge $q$ is allowed to move in one dimension between hard walls located at $x=\pm a$. The unperturbed system is described by the Hamiltonian
\begin{equation}\label{eq:H0-1Dp}
\hat{H}_{0} = \dfrac{\hbar^{2}\hat{p}^2}{2m} + \hat{V}(\hat{x}),
\end{equation}
with the potential
\begin{equation}
\hat{V}(\hat{x}) = \begin{cases}
0,&\quad |x|<a,\\
\infty, &\quad |x| \geqslant a.
\end{cases}
\end{equation}
The energy eigenvalues of this system are given by~\cite{sakuraiModernQuantumMechanics2017},
\begin{equation}\label{eq:1D-energy}
E_n = \frac{\hbar^2\pi^2}{8ma^2}n^2,\quad n\in\mathbb{N}.
\end{equation}
The particle is then subjected to a periodic driving of the form
\begin{equation}\label{eq:E-1Dp}
\bm{E}(t) = E_0\cos(\omega t)\hat{\bm{e}}_x.
\end{equation}
Let us view how to model the action of the external field in different gauges in practice.

\subsection{Electric dipole approximation: length and velocity gauges}

The electric dipole approximation can be employed when the wavelength of the radiation is much greater than the scale of the spatial variation of the system. It is then safe to assume that the external field does not depend on spatial degrees of freedom. Within this approximation, we consider the so-called
\emph{length} gauge~\cite{cohen-tannoudjiQuantumMechanics1977}:
\begin{equation}
\hat{\phi}(t) = -\bm{E}(t)\cdot\hat{\bm{r}}, \;\hat{\bm{A}}(t) = \hat{\bm{0}}.
\end{equation}
The Hamiltonian is then given by
\begin{equation}\label{eq:len_g_hamil}
\hat{H}^{E}(t) = \hat{H}_0 -q\bm{E}(t)\cdot\hat{\bm{r}},
\end{equation}
where the superscript $E$ refers to the explicit dependence on the electric field, following Ref.~\cite{passosNonlinearOpticalResponses2018}.

Consider now the gauge transformation
\begin{equation}
\hat{\Lambda}(\hat{\bm{r}}, t) = \bm{A}(t)\cdot\hat{\bm{r}}.
\end{equation}
Inserting the above and $\hat{H}^{E}(t)$ of Eq.~\eqref{eq:len_g_hamil} into the transformation relation of Eq.~\eqref{eq:hamil_trans_prop}, the Hamiltonian in the new gauge, that is commonly referred to as the \emph{velocity} gauge~\footnote{In the context of atomic physics, this gauge is more commonly referred as the radiation gauge. The Hamiltonian in the velocity gauge is given by $\hat{H}_0 - \frac{q}{m}\bm{A}(t)\cdot\hat{\bm{p}}$. Both gauges differ only by a time dependent operator multiple of the identity, which is irrelevant to the dynamics.}, becomes
\begin{equation}\label{eq:vel_g_hamil}
\hat{H}^{V}(t) = \text{exp}\left[\frac{iq}{\hbar}\bm{A}(t)\cdot\hat{\bm{r}}\right]\hat{H}_0\;\text{exp}\left[\frac{-iq}{\hbar}\bm{A}(t)\cdot\hat{\bm{r}}\right].
\end{equation}
In obtaining the above expression, the $d\hat{\Lambda}/{dt}$ term of Eq.~\eqref{eq:hamil_trans_prop} was canceled by the position operator term of Eq.~\eqref{eq:len_g_hamil}.

Employing the commutator identity $[\hat{r}_{i},\hat{p}_{j}]=i\hbar\delta_{ij}\hat{\mathds{1}}$, $\hat{H}^{V}(t)$ can be recast as
\begin{equation}\label{eq:vel_g_appr}
\hat{H}^{V_{2}}(t) = \hat{H}_0 - \frac{q}{m}\bm{A}(t)\cdot\hat{\bm{p}} + \frac{q^2A^2(t)}{2m}\hat{\mathds{1}},
\end{equation}
which is the more familiar expression usually found in textbooks~\cite{cohen-tannoudjiQuantumMechanics1977, scullyQuantumOptics1997}.

\subsubsection{Remark on the velocity gauge}

While correct in the ideal case, the Hamiltonian of Eq.~\eqref{eq:vel_g_appr} is only valid if one employs a complete set of bands and a Hamiltonian exactly expressible as Eq.~\eqref{eq:basic_hamil} for the calculation of its matrix elements; but this is hardly ever the case in practical calculations, where the set of basis states is usually truncated and therefore the form of $\hat{H}_0$ is no longer given by Eq.~\eqref{eq:basic_hamil}. A clarifying view on this issue has been provided recently in Refs.~\cite{passosNonlinearOpticalResponses2018, venturaGaugeCovariancesNonlinear2017}. The authors noted that, unlike in the length gauge, the form of the interaction term depends explicitly on $\hat{H}_0$ in the velocity gauge. As a consequence, a truncation of $\hat{H}_0$ implies a different form of the perturbation when using the expression in Eq.~\eqref{eq:vel_g_appr}, leading to results that are incorrect. The correct expression for the Hamiltonian in the velocity gauge, even for a truncated set of states, is provided by Eq.~\eqref{eq:vel_g_hamil}. In Appendix~\ref{app:piab_extra}, we further analyze this point in the context of the breakdown of the commutator identity $[\hat{r}_{i},\hat{p}_{j}]=i\hbar\delta_{ij}\hat{\mathds{1}}$.

\subsection{Quasienergy dispersion}

\subsubsection{Matrix elements and numerical details}

We now turn to the numerical calculation of the quasienergies in the various forms of the Hamiltonian. To provide a matrix representation, we consider the set of states that make the unperturbed Hamiltonian in Eq.~\eqref{eq:H0-1Dp} diagonal, $\hat{H}_0\ket{n} = E_n\ket{n}$. In this basis, the matrix elements of the Hamiltonian in the length gauge are given by
\begin{equation}\label{eq:len_m_hamil}
H^{E}_{nm}(t) = E_n\delta_{nm}-q\bm{E}(t)\cdot\bm{r}_{nm}.
\end{equation}
The above involves the position matrix elements $\bm{r}_{nm}$, which can be computed analytically for this particular system:
\begin{align}\label{eq:r_in_pot_well}
r_{nm} &= \begin{cases}
-\frac{16a}{\pi^2}\frac{nm}{(n^2-m^2)^2},&\quad n+m\;\text{odd},\\
0,&\quad n+m\;\text{even}.
\end{cases}
\end{align}
These matrix elements are also employed when computing the Hamiltonian matrix elements in the velocity gauge from Eq.~\eqref{eq:vel_g_hamil},
\begin{equation}\label{eq:vel_m_hamil}
H^{V}_{nm}(t) = \sum_{ls}W_{nl}\cdot E_{l}\delta_{ls}\cdot W^{\dagger}_{ms},
\end{equation}
with $(\hat{W})_{nm} = \left(\text{exp}\left[\frac{iq}{\hbar}\bm{A}(t)\cdot\hat{\bm{r}}\right]\right)_{nm}$. Finally, the matrix elements of the velocity gauge Hamiltonian in the form of Eq.~\eqref{eq:vel_g_appr} are given by
\begin{equation}\label{eq:vel_m_hamil_appr}
H^{V_{2}}_{nm}(t) = E_n\delta_{nm} - \frac{q}{m}\bm{A}(t)\cdot\bm{p}_{nm} +\dfrac{q^2A^2(t)}{2m}\delta_{nm},
\end{equation}
which involve the momentum matrix elements:
\begin{align}
p_{nm} &= \begin{cases}
-\frac{2i\hbar}{a}\frac{nm}{n^2-m^2},&\quad n+m\;\text{odd},\\
0,&\quad n+m\;\text{even} \end{cases}
\end{align}
In all cases, we calculate the quasienergy spectrum using a discretized expression of the one-cycle evolution operator; the technical details regarding the numerical scheme can be found in Appendix~\ref{app:numerical}.

\subsubsection{Results}

\begin{figure}[!tp]
\centering
\includegraphics[width=\columnwidth]{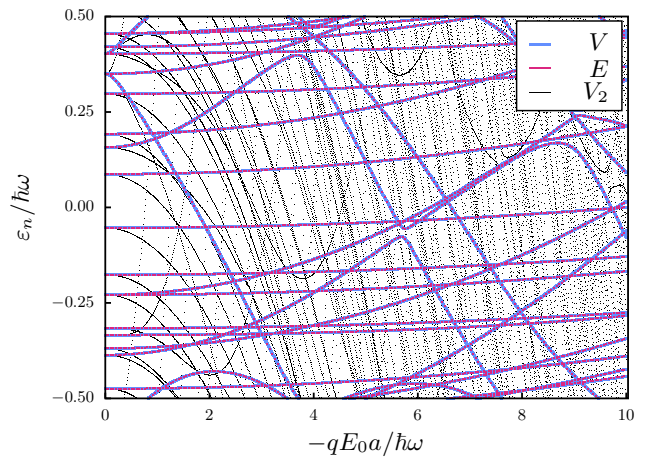}
\caption{Quasienergy spectrum of the first 20 states in the first quasienergy Brillouin zone of a system composed by a particle of charge $q$ and mass $m$ in an infinite potential well of width $a$ subjected to a driving given by the electrical field $E(t) = E_0\cos(\omega t)$. The frequency $\omega$ is fixed such that $\hbar\omega = 0.95\times 3\hbar^2\pi^2/(2ma^2)$ and $E_0$ is varied in the $[0, -10\hbar\omega/qa]$ range. The red line shows the spectrum calculated using the length gauge matrix elements in Eq.~\eqref{eq:len_m_hamil}, $\varepsilon_n^E$. The blue and black lines show the spectrum calculated using the velocity gauge matrix elements in the form of Eqs.~\eqref{eq:vel_m_hamil} and~\eqref{eq:vel_m_hamil_appr}, $\varepsilon_n^V$ and $\varepsilon_n^{V_2}$, respectively.}
\label{fig:one_D_quasienergy_spectrum}
\end{figure}

Figure.~\ref{fig:one_D_quasienergy_spectrum} shows the calculated quasienergy spectrum of the 20 lowest eigenstates as a function of the external field magnitude in the range $E_0\in[0, -10\hbar\omega/qa]$. We have fixed the frequency $\omega$ such that the corresponding energy is 5\% red-detuned from the first allowed transition, that is
\begin{equation}\label{eq:resonance}
\hbar\omega = 0.95\times(E_2-E_1).
\end{equation}
This is precisely the case considered in Ref.~\cite{holthausFloquetEngineeringQuasienergy2016}, where the quasienergy spectrum was calculated using the length gauge.

The figure shows three different calculations corresponding to ${H}^{E}_{nm}(t)$, ${H}^{V}_{nm}(t)$, and ${H}^{V_{2}}_{nm}(t)$. The quasi-energy dispersion for ${H}^{E}_{nm}(t)$ and ${H}^{V}_{nm}(t)$ agree up to numerical precision in all the considered driving field ranges and for all eigenstates. This offers the numerical verification we were looking for, namely the gauge-invariant nature of quasienergies as a result of the gauge-covariance of the Floquet Hamiltonian derived in Eq.~\eqref{eq:eff_floq_hamil_trans_prop}. In addition, our results agree completely with the spectrum reported in Fig.~3 of Ref.~\cite{holthausFloquetEngineeringQuasienergy2016} up to an irrelevant rigid shift of the whole set of quasienergies. In particular, Fig.~\ref{fig:one_D_quasienergy_spectrum} correctly describes the strong dispersion of the two quasienergies located at $\varepsilon/\hbar\omega\simeq 0.35$ for low fields, which correspond to the lowest two eigenstates that are close to resonance in Eq.~\eqref{eq:resonance}~\cite{holthausFloquetEngineeringQuasienergy2016}.

On the other hand, the quasienergy spectrum corresponding to $\hat{H}^{V_{2}}_{nm}(t)$ in Fig.~\ref{fig:one_D_quasienergy_spectrum} completely deviates from the other two curves and features a nonphysical behavior; only in the weak field limit $E_0\rightarrow 0$ does it approach the correct value. This offers a clear numerical evidence of the inadequacy of the velocity gauge expression in Eq.~\eqref{eq:vel_m_hamil} for performing calculations using a finite basis. We have verified that this deviation decreases as the number of eigenstates in the calculation increases, as expected. In Appendix~\ref{app:piab_extra} we study an improved version of the $\hat{H}^{V_{2}}(t)$ Hamiltonian obtained by adding an expansion that is introduced in the coming section.

\section{Extended systems}\label{sec:extended_systems}

The particle in a box described in the previous section constitutes an example of a finite system; atoms and molecules are real-life examples of this type. In contrast, electrons in a crystal or atoms trapped in a periodic optical lattice constitute an {extended} system. In these cases, the potential entering the ground state Hamiltonian, Eq.~\eqref{eq:basic_hamil} generally obeys the periodic property
\begin{equation}
\hat{V}(\hat{\bm{r}})=\hat{V}(\hat{\bm{r}}+\bm{R}\hat{\mathds{1}}),
\end{equation}
where $\bm{R}$ is a Bravais lattice vector. When subjected to a time dependent perturbation, the calculation of the Hamiltonian in this type of system requires extra care. The fundamental reason for the additional difficulty lies in the position operator, whose representation becomes subtle in an extended system~\cite{blountFormalismsBandTheory1962}. In this section we describe the scheme we adopted to deal with $\hat{\bm{r}}$ in practical Floquet calculations in extended systems.

\subsection{Quantum-geometric gauge and representation of \texorpdfstring{$\hat{\bm{r}}$}{r}}

Bloch's theorem states that the eigenstates of the periodic Hamiltonian $\hat{H}_0$ are of the form
\begin{equation}
\ket{\psi_n(\bm{k})} = \text{exp}\left[i\bm{k}\cdot\hat{\bm{r}}\right]\ket{u_{n}(\bm{k})},
\end{equation}
with
\begin{equation}\label{eq:def_bloch_states}
\begin{split}
&\braket{\bm{r}|\psi_n(\bm{k})} = \psi_{n\bm{k}}(\bm{r}) = e^{i\bm{k}\cdot\bm{r}}u_{n\bm{k}}(\bm{r}),\\
&\;u_{n\bm{k}}(\bm{r} + \bm{R}) = u_{n\bm{k}}(\bm{r}).
\end{split}
\end{equation}
This basis transforms as irreducible representations of the Bravais lattice translation operators $\hat{\mathcal{T}}(\bm{R})$, and its labeling corresponds to a vector $\bm{k}$ within the first Brillouin zone (BZ) of the system~\cite{evarestovSiteSymmetryCrystals1997}. In this extended basis, the closure and orthogonality properties read
\begin{subequations}\label{eq:orth-closure}
\begin{align}
&\hat{\mathds{1}} = \int_{\text{BZ}}\frac{d^dk}{(2\pi)^d}\sum_{n=1}^{\infty}\ket{\psi_n(\bm{k})}\bra{\psi_n(\bm{k})}, \label{eq:extended_resolution_of_identity}\\
&\braket{\psi_n(\bm{k})|\psi_m(\bm{k}')} = (2\pi)^d\delta_{nm}\delta(\bm{k}-\bm{k}'),
\end{align}
\end{subequations}
respectively, where $d$ is the dimensionality of the system.

There is a gauge freedom in the definition of Bloch functions; we shall refer to it as the ``quantum-geometric" gauge freedom, to differentiate it from the electromagnetic one discussed in Secs.~\ref{sec:em-gauge} and~\ref{sec:particle_in_a_box}. Consider a set of $N$ eigenstates of $\hat{H}_0$. Then, at each wavevector $\bm{k}$, a unitary rotation $\hat{\mathcal{U}}$ among the states can be performed
\begin{equation}\label{eq:quantum_geometric_gauge_freedom}
\ket{\psi_m'(\bm{k})} = \sum_{n=1}^{N}\mathcal{U}_{nm}(\bm{k})\ket{\psi_n(\bm{k})},
\end{equation}
such that the orthogonality and closure properties in Eq.~\eqref{eq:orth-closure} are left invariant~\cite{bohmGeometricPhaseQuantum2003}. In the special case where $\hat{\mathcal{U}}$ is chosen diagonal, ${\mathcal{U}}_{nm}(\bm{k})=\delta_{nm}e^{i\phi_{n}(\bm{k})}$, the rotated Bloch functions remain eigenstates of the Hamiltonian, but this need not be the case for a general unitary rotation that mixes different bands.

In analogy with the discussion in Sec.~\ref{sec:gauge-covariance} for the electromagnetic gauge transformation $\hat{\Lambda}(t)$, some operators transform gauge-covariantly under $\hat{\mathcal{U}}$ whereas others do not. The Hamiltonian $\hat{H}_0$, for instance, is gauge-covariant under $\hat{\mathcal{U}}$, namely $H'_0(\bm{k}) = \mathcal{U}^{\dagger}(\bm{k})H_0(\bm{k})\mathcal{U}(\bm{k})$ where each operator is represented by an $N\times N$ matrix~\cite{wangInitioCalculationAnomalous2006}. On the other hand, consider the matrix elements representing the Berry connection, a quantity that plays a central role in the discussion of quantum-geometric properties of solids~\cite{vanderbiltBerryPhasesElectronic2018}:
\begin{equation}\label{eq:berry_conn}
\xi^j_{nm}(\bm{k}) = i\braket{u_{n}(\bm{k})|\frac{\partial}{\partial k_j}|u_{m}(\bm{k})}.\end{equation}
Because of the presence of the derivative in its definition, the transformation under $\hat{\mathcal{U}}$ acquires an extra piece~\cite{wangInitioCalculationAnomalous2006},
\begin{equation}\label{eq:xi}
\xi'^j(\bm{k}) = \mathcal{U}^{\dagger}(\bm{k})\xi^j(\bm{k})\mathcal{U}(\bm{k}) + i\frac{\partial \mathcal{U}(\bm{k})}{\partial k_j}\mathcal{U}^{\dagger}(\bm{k}),
\end{equation}
and therefore does not transform covariantly in general.

Let us now come to the position operator. One can show (see Appendix~\ref{app:resolution_of_r}) that, in the Bloch representation, the operator can be expressed as~\cite{blountFormalismsBandTheory1962}
\begin{equation}\label{eq:res_of_r_in_r_basis}
\hat{\bm{r}} = \int_{\text{BZ}}\int_{\text{BZ}}\frac{d^dk\;d^dq}{(2\pi)^d}\sum_{n,m=1}^{\infty}\left(i\bm{\mathcal{D}}\right)_{nm}\ket{\psi_m(\bm{q})}\bra{\psi_n(\bm{k})},
\end{equation}
where $\hat{\bm{\mathcal{D}}}$ is the quantum-geometric covariant derivative~\cite{venturaGaugeCovariancesNonlinear2017,parkerDiagrammaticApproachNonlinear2019}
\begin{equation}\label{eq:covder}
(\mathcal{D}^j)_{nm} = -\delta_{nm}\partial_{k_j}\delta(\bm{k}-\bm{q}) - i\xi^{j}_{nm}(\bm{k})\delta(\bm{k}-\bm{q}).
\end{equation}
As implied by its name, this object transforms covariantly under $\hat{\mathcal{U}}$, even if the individual pieces that conform it do not [c.f.~Eq.~\eqref{eq:xi}]. Therefore, finding a matrix representation of $\hat{\bm{r}}$ reduces to computing the matrix elements of Eq.~\eqref{eq:covder}.

In the language of differential geometry~\cite{bohmGeometricPhaseQuantum2003}, we are analyzing a fiber bundle whose base manifold is the $d$-torus (the BZ), the typical fiber is the $N$-dimensional Hilbert space where the $\ket{u_n(\bm{k})}$ are defined and the gauge group is $U(N)$. For each variable $k_j$ parametrizing the base manifold, there exists an associated covariant derivative $(i\hat{\mathcal{D}}^j)$ and connection form $\hat{\xi}^j(\bm{k})$. These objects are required to provide a notion of \emph{parallel transportation} and are solely defined though their respective transformation properties.

\subsection{Perturbative expansion in the velocity gauge}\label{sec:ext_sys_int_hamil}

The above considerations need to be taken into account for calculating Floquet quasienergies in extended systems through the Hamiltonian $\hat{H}(t)$ in either form, length gauge [Eq.~\eqref{eq:len_m_hamil}], or velocity gauge [Eq.~\eqref{eq:vel_m_hamil}]. The most challenging aspect is numerical implementation: taking the derivative of the delta function in Eq.~\eqref{eq:covder} is problematic since it involves the nonlocal difference $\bm{k}-\bm{q}$. It is worth noting that the difficulty faced here is more severe than that encountered in the perturbative analysis of optical responses of solids. In the latter case and taking as reference the length gauge, $\hat{\bm{r}}$ always enters the response coefficients via commutators with an operator $\hat{O}$, which is the well-defined covariant derivative of operator $\hat{O}$~\cite{aversaNonlinearOpticalSusceptibilities1995,parkerDiagrammaticApproachNonlinear2019, passosNonlinearOpticalResponses2018}
\begin{equation}
\begin{split}
&\hat{\mathcal{D}^j}[\hat{O}]\equiv[\hat{\mathcal{D}^j},\hat{O}] = \int_{\text{BZ}}\frac{d^dk}{(2\pi)^d}\sum_{n,m=1}^{\infty}\times\\
&\left[\frac{\partial O_{nm}(\bm{k})}{\partial k_j} -i[\xi^j(\bm{k}), O(\bm{k})]_{nm}\right]\ket{\psi_n(\bm{k})}\bra{\psi_m(\bm{k})}.\label{eq:covder_action_on_operator}
\end{split}
\end{equation}
In the case that concerns us, on the other hand, calculating $\hat{H}(t)$ via the length gauge expression of Eq.~\eqref{eq:len_m_hamil} requires computing the position matrix element plainly. Previous works~\cite{holthausQuantumTheoryIdeal1992, holthausCollapseMinibandsFarinfrared1992, honeLocallyDisorderedLattices1993, holthausFloquetEngineeringQuasienergy2016} have proceeded by considering just the interband piece of the Berry connection, namely by approximating
\begin{equation}
\label{eq:r-app}
{\bm{r}}_{nm}(\bm{k}) \simeq (1-\delta_{nm})\bm{\xi}_{nm}(\bm{k}).
\end{equation}
While this has the obvious advantage of not having to deal with the Dirac delta piece in Eq.~\eqref{eq:covder}, it appears to have several drawbacks. In the first place, information of the intraband contribution to the Berry connection (proportional to $\delta_{nn}$) is completely lost. As for the discarding of the Dirac delta piece, it seems a rather uncontrolled procedure considering the highly singular nature of this contribution. Finally on a more fundamental level, Eq.~\eqref{eq:r-app} maintains gauge-covariance only under a diagonal transformation of the form ${\mathcal{U}}_{nm}(\bm{k})=\delta_{nm}e^{i\phi_{n}(\bm{k})}$ [gauge group $U(1)$], but it is not gauge-covariant under a general transformation $\hat{\mathcal{U}}$ [gauge group $U(N)$].

In view of this, we have opted to work in the velocity gauge as it allows a perturbative treatment that can be truncated in a controllable fashion. By replacing $\hat{\bm{r}}\rightarrow i\hat{\bm{\mathcal{D}}}$ and using the Baker-Campbell-Hausdoff lemma~\cite{sakuraiModernQuantumMechanics2017} in Eq.~\eqref{eq:vel_g_hamil} one gets~\cite{parkerDiagrammaticApproachNonlinear2019, passosNonlinearOpticalResponses2018}
\begin{equation}\label{eq:vel_g_in_ext_systems}
\hat{H}(t) = \lim_{M\to \infty}\sum_{n=0}^{M}\frac{1}{n!}\frac{(-q)^n}{\hbar^n}\prod_{j=1}^{n}\sum_{\alpha_j = 1}^d A_{\alpha_j}(t) \hat{\mathcal{D}}^{\alpha_j}[\hat{H}_0],
\end{equation}
which is a power series of objects involving the successive application of the quantum-geometric covariant derivative of Eq.~\eqref{eq:covder_action_on_operator} on $\hat{H}_0$. In a practical calculation one needs to consider a finite number $M$ of expansion terms. Their explicit element wise expression can be straightforwardly obtained using Eq.~\eqref{eq:covder_action_on_operator}. In this work we evaluate the above terms by employing the so-called Wannier interpolation technique; a description of the gauge choice and further technical details are given in Appendix~\ref{app:wannier}.

\subsubsection{Quantifying truncation errors}\label{sec:conv_of_series}

Equation~\eqref{eq:vel_g_in_ext_systems} is a power series expansion on the external field. Let us consider that the action of the covariant derivative on $\hat{H}_0$ contributes an amount that is of the order of some characteristic length scale $l$. Then, each term in the expansion introduces the multiplicative factor
\begin{equation}\label{eq:lambda}
\frac{\lambda}{n} = \frac{-qEl}{n\hbar\omega},
\end{equation}
acting on $\hat{H}_0$, which is proportional to the electric field amplitude $E$ and the length scale $l$, and inversely proportional to the characteristic driving frequency $\omega$, owing to $A\sim E/\omega$. The parameter $\lambda$ defined above is helpful in classifying different driving regimes.

Working with the above assumption, the driving inducing operator $\text{exp}\left[\frac{iq}{\hbar}\bm{A}(t)\cdot\hat{\bm{r}}\right]$ can be approximated by $e^{\lambda}$. Then, we can define a proxy to quantify the relative error made by keeping $M$ terms in the series Eq.~\eqref{eq:vel_g_in_ext_systems},
\begin{equation}
err_M(\lambda) = 1 - \dfrac{\sum_{n=0}^M \frac{1}{n!}\lambda^n}{e^{\lambda}}.
\end{equation}
\begin{figure}[!tp]
\centering
\includegraphics[width=\columnwidth]{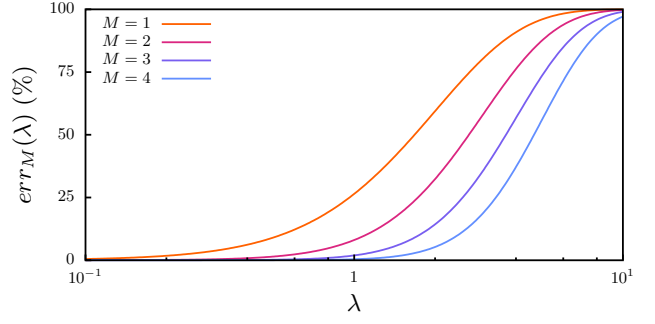}
\caption{Proxy to the relative error made by keeping only $M$ terms in the series expansion of the driving inducing operator $\text{exp}\left[\frac{iq}{\hbar}\bm{A}(t)\cdot\hat{\bm{r}}\right]$ for logarithmically spaced values of $\lambda$.}
\label{fig:error_estimates}
\end{figure}
In Fig.~\ref{fig:error_estimates} we illustrate the error estimates up to $M=4$. Let us first focus on $\lambda<1$, where central dynamical effects such as multiphoton resonances are expected to take place~\cite{holthausFloquetEngineeringQuasienergy2016}. In this range, the $M=2$ approximation already yields an error below $\sim10\%$, while for $M>2$ the error goes below $\sim1\%$. Therefore, it appears reasonable to expect that keeping the first few terms in the expansion of Eq.~\eqref{eq:vel_g_in_ext_systems} suffices to provide a fair estimate of this important regime. For $\lambda>1$, on the other hand, the error in Fig.~\ref{fig:error_estimates} quickly increases above $\sim10\%$ even for $M=4$. This regime corresponds to the large electric field or slow driving frequency limit.

\section{Graphitic BC\texorpdfstring{\textsubscript{2}}{\texttwoinferior}N}\label{sec:real_materials}

In this section we present the results of the calculations of Floquet quasienergies using the extended system approach of Sec.~\ref{sec:extended_systems} for the case of the material BC$_2$N.

\subsection{Technical details}

BC$_2$N is a layered semiconductor composed of alternating zigzag chains of carbon and boron nitride. Among its several polytypes, we consider the most stable noncentrosymmetric structure identified in Ref.~\cite{panInterlayerStackingNature2006}, namely the A2. It belongs to the Pmm2 space group (No. 25, unconventional setting 6, $(a, -c, b)$), with an orthorhombic lattice with constants $a = 2.47\;\text{\r{A}}$, $b = 4.32\;\text{\r{A}}$, $c = 6.44\;\text{\r{A}}$. As analyzed in detail in Ref.~\cite{ibanez-azpirozDirectionalShiftCurrent2020}, the system is left invariant under the mirror operation $\mathcal{M}_x$, which will be important in the following discussion.

We have conducted an \textit{ab-initio} analysis for this material. The first step consisted of a density functional theory calculation using the \textsc{Quantum Espresso}~\cite{giannozziUantumESPRESSOExascale2020} package. We employed a $10\times 10\times 10$ k-mesh both for the self-consistent calculation of the ground state as well as for the subsequent non-self-consistent calculation of the energy eigenvalues and Bloch eigenfunctions. The core-valence interaction was treated by means of scalar relativistic projector-augmented-wave pseudopotentials that had been generated with the Perdew-Burke-Ernzerhof exchange-correlation functional. The pseudopotentials were taken from PS library, generated using the \textsc{atomic} code by A. Dal Corso v.6.3, and the energy cutoff for the plane-wave basis expansion was set at 70 Ry. In the next step, we constructed maximally localized Wannier functions using the \textsc{Wannier90} package~\cite{pizziWannier90CommunityCode2020}. After discarding the two lowest-lying bands, we considered a set of eight disentangled Wannier functions around the Fermi level to span the ``inner" energy range~\cite{souzaMaximallyLocalizedWannier2001} $[-3.33, 3.35]\;\text{eV}$. As initial projections, we considered $p_z$ atom-centered orbitals for all atoms.

In order to calculate the quasienergy spectrum $\varepsilon_n(\bm{k})$, we implement the time discretization method detailed in Appendix~\ref{app:numerical}, with matrix representation of the Hamiltonian $\hat{H}(t_j)$ given by the different methods discussed in Sec.~\ref{sec:ext_sys_int_hamil}: the length gauge, Eq.~\eqref{eq:r-app}, and the velocity gauge Eq.~\eqref{eq:vel_g_in_ext_systems} keeping terms up to $M=1$, $M=2$ and $M=3$. We have performed a time complexity analysis of the algorithm in Appendix~\ref{app:time_complexity}, where we have studied the computational cost of the different methods presented in the manuscript.

\subsection{Band structure}

\begin{figure}[!tp]
\centering
\includegraphics[width=\columnwidth]{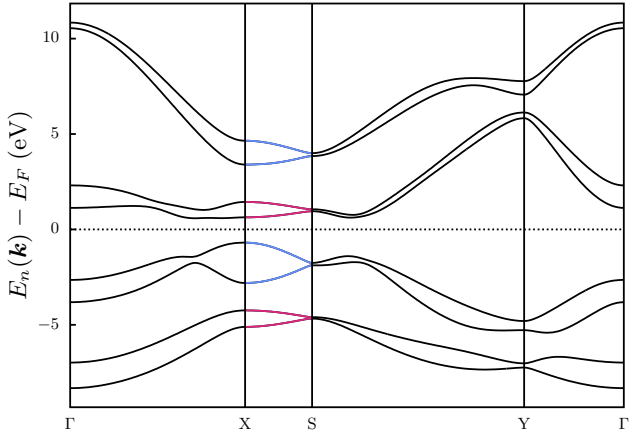}
\caption{Band structure of mirror-symmetric BC$_2$N along the $k_z = 0$ $\Gamma - \text{X} - \text{S} - \text{Y} - \Gamma$ path in the BZ and in the $[-9.0, 11.5]\;\text{eV}$ energy range centered around the Fermi energy. The displayed bands have a definite parity with respect to the $\mathcal{M}_x$ point group operation, $A_n$, which is displayed as a colored over line. The red (blue) overline represents the $+1$ ($-1$) eigenvalue. }
\label{fig:BC2N_band_structure}
\end{figure}

We show the band structure of the last four valence and first four conduction bands of the system along the $k_z = 0$ plane in Fig.~\ref{fig:BC2N_band_structure}. These bands are expected to account for most of the optical transitions that will be discussed.
The figure shows that the bands are packed in pairs, with an almost constant energy difference between pair elements along the displayed path. The bands in the $\text{X}-\text{S}$ path have definite parity with respect to the $\mathcal{M}_x$ point group operation, which we denote as $A_n$. The eigenvalue with respect to this symmetry operation ($\pm 1$) is denoted by a red over line for the eigenvalue $+1$ and a blue over line for the eigenvalue $-1$.

\subsection{Quasienergy dispersion}

\subsubsection{General properties and quality of approximations}\label{sec:quasienergy_dispersion}

We study the case where the system is subjected to a periodic driving given by the linearly polarized electrical field
\begin{equation}\label{eq:driving}
\bm{E}(t) = E\cos(\omega t)\hat{e}_x,
\end{equation}
and we consider the dipole approximation.
\begin{figure}[!tp]
\centering
\includegraphics[width=\columnwidth]{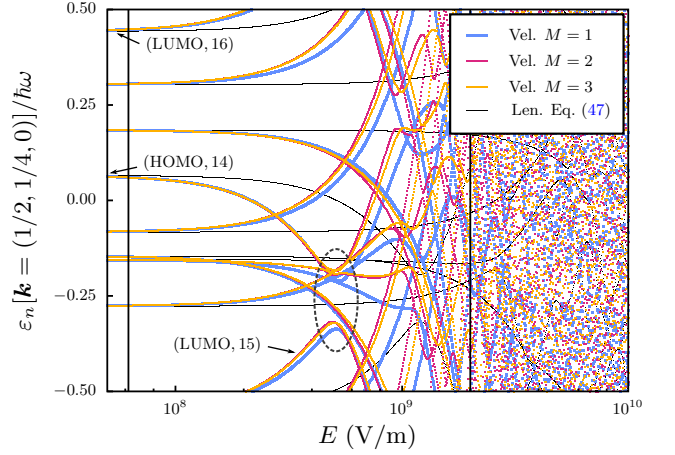}
\caption{Quasienergy spectrum of the last four valence bands and first four conduction bands of mirror-symmetric BC$_2$N in the BZ point $\bm{k} = (1/2, 1/4, 0)$ in crystal coordinates. The spectra have been calculated with the Hamiltonian given in four different approximations: keeping terms up to $M=1$ (blue), $M=2$ (red), $M=3$ (yellow) in the series Eq.~\eqref{eq:vel_g_in_ext_systems} and calculation done by employing the approximation Eq.~\eqref{eq:r-app} (black). The driving field is given by $\bm{E}(t) = E\cos(\omega t)\hat{e}_x$, where $E$ is varied and $\hbar\omega = 0.5\;\text{eV}$. The tuples $(n, m)$ near the y axis of figures denote the corresponding band index for the HOMO ($n=4$) and LUMO ($n=5$) states and quasienergy representative in the $E\rightarrow 0$ limit, respectively, such that $\lim_{E\to 0}\varepsilon_n(\bm{k}) = E_n(\bm{k}) - m \hbar\omega$. The noninteracting, intermediate and continuum regimes are separated by vertical lines. The dashed oval denotes a single photon resonance between HOMO and LUMO states. In the implementation of Eq.~\eqref{eq:numerical_one_per_tev}, the period $T = 2\pi/\omega$ has been discretized in 513 steps.}
\label{fig:BC2N_quasienergies_amp}
\end{figure}
Figure~\ref{fig:BC2N_quasienergies_amp} shows the quasienergy spectrum of the eight bands considered in Fig.~\ref{fig:BC2N_band_structure} as a function of the electric field amplitude. The calculations were performed at k-point $\bm{k} = (1/2, 1/4, 0)$ (crystal coordinates) lying along the X-S line, and fixed frequency $\hbar\omega = 0.5\;\text{eV}$, which corresponds to an almost resonant excitation between the highest occupied molecular orbital (HOMO) and the lowest unoccupied molecular orbital (LUMO). The figure includes three results calculated in the velocity gauge obtained by keeping a different number of expansion terms in Eq.~\eqref{eq:vel_g_in_ext_systems}; up to $M=1$, $M=2$, and $M=3$. In addition, we have also included results calculated using the length gauge approximation of Eq.~\eqref{eq:r-app}, employed in other works.

Figure~\ref{fig:BC2N_quasienergies_amp} distinguishes three regimes as a function of $E$, which we have separated by vertical lines for the case of the velocity gauge results as a guide to the eye. The first is the noninteracting regime of small $E$, where quasienergy bands are virtually flat. The second is an intermediate electric field regime where the onset of dispersion takes place and is the physically most interesting one. Among other features, this regime shows wide, pronounced avoided crossings that mark multiphoton resonances, which we study in detail in Sec.~\ref{sec:generalized_parity} below. The third regime is the continuum limit of large $E$, which corresponds to $\lambda\gg 1$ [c.f. Eq.~\eqref{eq:lambda}] wherein transitions between virtually all energy levels are possible, yielding a featureless continuum spectrum.

We turn next to analyze the different levels of approximation. The figure shows a rather good agreement between the velocity gauge approximations; the change in the main spectral features like the avoided crossing between the HOMO and LUMO states at $E\approx 5\times 10^8\;\text{V/m}$, is well below $10\%$. According to the discussion of Sec.~\ref{sec:conv_of_series} and Fig.~\ref{fig:error_estimates}, the error with respect to the exact result is expected to be below $1\;\%$ for the calculation with $M=3$. In contrast, the length gauge result clearly deviates from the velocity gauge results and overestimates the value of $E$ for the onset of dispersion and the spectral features therein, in some cases by more than an order of magnitude. This exemplifies the uncontrolled nature of the approximation of Eq.~\eqref{eq:r-app}, where the contribution to the quasienergy of the discarded terms is not guaranteed to be small.

\begin{figure*}[htpb]
\centering
\includegraphics[width=2\columnwidth]{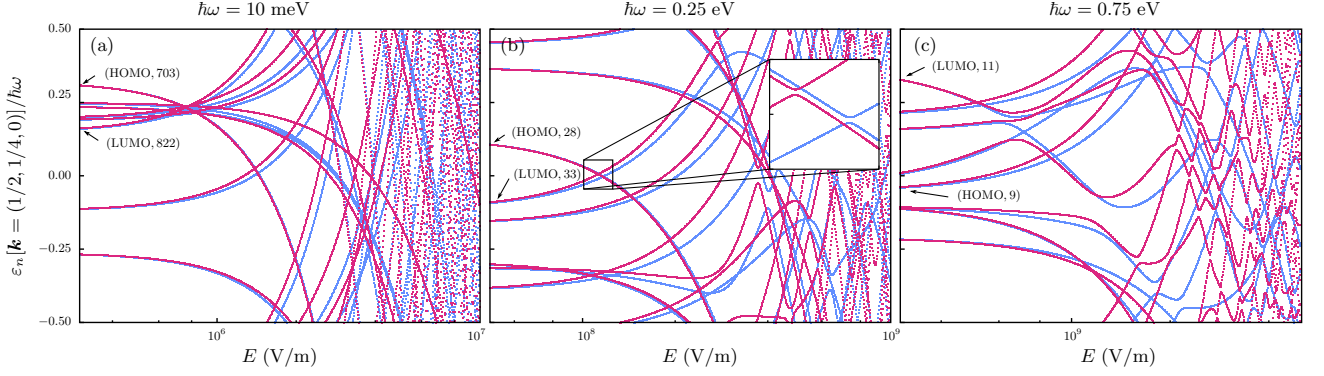}
\caption{Quasienergy spectrum of the last four valence bands and first four conduction bands of mirror-symmetric BC$_2$N in the BZ point $\bm{k} = (1/2, 1/4, 0)$ in crystal coordinates. The spectra have been calculated with the Hamiltonian in two different approximations: keeping terms up to $M=1$ (blue), and $M=2$ (red) in Eq.~\eqref{eq:vel_g_in_ext_systems}. The driving field is given by $\bm{E}(t) = E\cos(\omega t)\hat{e}_x$, where $E$ is varied and $\hbar\omega = 10\;\text{meV}$ (a), $0.25\;\text{eV}$ (b), and $0.75\;\text{eV}$ (c). In (b), we highlight the correction of an avoided crossing between HOMO and LUMO states. In the implementation of Eq.~\eqref{eq:numerical_one_per_tev}, the period $T = 2\pi/\omega$ has been discretized in 513 steps.}
\label{fig:BC2N_quasienergies}
\end{figure*}

In Fig.~\ref{fig:BC2N_quasienergies} we inspect the dependence of the quasienergy spectrum on the driving-specific frequency for selected values $\hbar\omega = 10\;\text{meV}$ (a), $0.25\;\text{eV}$ (b), and $0.75\;\text{eV}$ (c); we only show results calculated in the velocity gauge since the length gauge approximation of Eq.~\eqref{eq:r-app} differs by more than an order of magnitude as in the case depicted in Fig.~\ref{fig:BC2N_quasienergies_amp}. The displayed features indicate that the electric field required to trigger the onset of interactive dynamics and the continuum limit roughly scale with $\omega$, in accordance with the definition of the $\lambda$ parameter in Eq.~\eqref{eq:lambda}. Furthermore, all frequencies show an acceptable agreement between the two velocity gauge approximations up until the continuum limit. This implies that the accuracy of a given approximation to the series in Eq.~\eqref{eq:vel_g_in_ext_systems} does not appear to depend strongly on the driving-specific parameters.

\subsubsection{Avoided crossings, multiphoton resonances, and generalized parity}\label{sec:generalized_parity}

To conclude the analysis of our results we study in more detail the phenomenon of avoided crossings, which signal a strong coupling between the Floquet states involved~\cite{holthausQuantumTheoryIdeal1992,holthausFloquetEngineeringQuasienergy2016}. As reviewed in Sec.~\ref{sec:floq}, a Floquet quasienergy [c.f. Eq.~\eqref{eq:quasienergy}] is defined by the two integers $n$ and $m$ that denote the band index and quasienergy representative in the $E\rightarrow 0$ limit, respectively:
\begin{equation}\label{eq:traceback_definition}
\lim_{E\to 0}\varepsilon_n(\bm{k}) = E_n(\bm{k}) - m \hbar\omega.
\end{equation}
Using the wording of Ref.~\cite{holthausQuantumTheoryIdeal1992}, the index $m$ represents ``how many photons have to be subtracted from $E_n$ in order to arrive in the quasienergy first Brillouin zone".

Employing group theory arguments~\cite{evarestovSiteSymmetryCrystals1997}, it can be checked that if the eigenvalues at a given $\bm{k}$ have a definite parity under a improper symmetry operation, such as a mirror $\mathcal{M}_x$, then these also have a definite parity under the combined operation
\begin{equation}
P(\bm{k}, t) = \begin{cases}
&k_x\to -k_x,\\
&t\to t + T/2.
\end{cases}
\end{equation}
The Wigner-von Neumann non-crossing theorem~\cite{vonneumanUberMerkwurdigeDiskrete1929} states that if only one parameter is varied, in this case the electric field amplitude $E$, then states with the same parity cannot cross. Then, the avoided crossing between pairs $(n, m)$ and $(n', m')$ mark a multiphoton resonance between bands $n, n'$ with energy $\hbar\omega|m-m'|$.

For conciseness, let us exemplify these notions by focusing on the avoided crossing at $E\approx 5\times 10^8\;\text{V/m}$ between the HOMO and the LUMO states highlighted in Fig.~\ref{fig:BC2N_quasienergies_amp}. Since $\bm{k}$ lies along the $\text{X}-\text{S}$ line for this case, the quasienergies have a definite parity under $P$ given by~\cite{holthausQuantumTheoryIdeal1992, holthausFloquetEngineeringQuasienergy2016},
\begin{equation}
P_{nm} = A_n(-1)^{m+1},
\end{equation}
where $A_n$ is the parity with respect to the $\mathcal{M}_x$ operation, as introduced in the discussion of the band structure in Fig.~\ref{fig:BC2N_band_structure}. In the case of the HOMO and LUMO states we have $A_{\text{HOMO}} = -A_{\text{LUMO}} = -1$, while the representatives are $m = 14$ for the HOMO and $m^{\prime} = 15$ for the LUMO state (notice that translation between quasienergy first Brillouin zone is accompanied by a shift in the representative in Fig.~\ref{fig:BC2N_quasienergies_amp}), hence $P_{\text{HOMO},14}=P_{\text{LUMO},15}$. From the group theory perspective, the quasienergies corresponding to these states cannot cross and therefore undergo an avoided crossing that marks a one-photon resonance since $|m-m^{\prime}|=1$.

Figure~\ref{fig:BC2N_quasienergies} displays the representative of the HOMO and LUMO states for varying frequency; as expected, it decreases with increasing $\omega$. Following the same line of argument as before, one concludes that for $\hbar\omega=10\;\text{meV}$ [Fig.~\ref{fig:BC2N_quasienergies}(a)] and $\hbar\omega=0.25\;\text{eV}$ [Fig.~\ref{fig:BC2N_quasienergies}(b)] the states cannot cross given that $P_{\text{HOMO},m}=P_{\text{LUMO},m^{\prime}}$ in both cases (see representative indices in the left hand side of the figures). As expected, the smaller $\omega$ is, the more photons intervene in the multiphoton resonances; we observe a five-photon resonance for $\hbar\omega=0.25\;\text{eV}$ at $E\approx 1.1\times 10^8\;\text{V/m}$ (see inset), while for $\omega=10$ meV the avoided crossing takes place at $E\approx 8.0\times 10^5\;\text{V/m}$ marking a 119-photon process, but it is too narrow to visualize. Note that the values of the electric field needed to trigger these multi-photon resonances are in the range of experimentally attainable values. For $\hbar \omega=0.75\;\text{eV}$ on the other hand, Fig.~\ref{fig:BC2N_quasienergies}(c) displays no direct interaction between the HOMO and LUMO states given that $P_{\text{HOMO},9}=-P_{\text{LUMO},11}$, implying that the non-crossing theorem does not apply. This could be explained by considering response theory arguments, since the driving frequency is larger than the band gap at $\bm{k} = (1/2, 1/4, 0)$, the states cannot be coupled by light.

\section{Summary and Outlook}

In this work, we have studied two central quantities of the Floquet theory, namely the effective Floquet Hamiltonian and its eigenvalues, the quasienergies. In the first part of the work, we have analyzed the implications of the electromagnetic gauge freedom in the transformation properties of these quantities. This subject has attracted relatively little attention in the literature so far, and it appears to have been only studied using perturbation theory up to second order in the external field~\cite{shtoffFloquetGaugeinvariantCoupled2003, sindelkaFloquetPerturbationTheory2007}. Here we have explicitly proved the gauge-covariance of the effective Floquet Hamiltonian, which ensures the gauge-invariance of the quasienergy spectrum to all orders. We have provided a numerical example of the latter by discussing the results obtained in the velocity and length gauges for a particle trapped in a one-dimensional box subjected to a periodic drive.

The second part of the analysis has been devoted to the methodology and implementation in extended systems. To this end, we have carefully considered the resolution of the position operator, whose interpretation corresponds to the quantum-geometric covariant derivative in this context. We have found it convenient to work in the velocity gauge, where the Hamiltonian can be recast as a power series of objects involving the applied electric field and the successive application of the covariant derivative on $\hat{H}_0$. This expansion can be truncated at the desired order, hence the Hamiltonian can be approximated in a controlled way. Furthermore, we have shown that all objects involved in the formalism can be computed efficiently by employing the so-called Wannier-interpolation technique, rendering high accuracy to the approach. We have demonstrated the method's capability by reporting the quasienergy spectrum of the semiconducting material BC$_2$N, where we have identified multiphoton resonances at experimentally attainable electric field and drive frequency values.

We expect the method to be useful in the computation of Floquet quasienergies, specially in cold atomic gases and real materials, given the adequacy of Wannier functions for describing extended systems. When needed, the precision of the results can be improved straightforwardly by including further terms in the expansion of the Hamiltonian of Eq.~\eqref{eq:vel_g_in_ext_systems}. Additional physical effects not analyzed in the present manuscript, such as dynamical localization in momentum space~\cite{holthausCollapseMinibandsFarinfrared1992} or the emergence of Floquet-Bloch bands in topological insulators~\cite{itoBuildupDephasingFloquet2023}, are natural targets for future studies.

\section{Acknowledgments}

This project has received funding from the European Union’s Horizon 2020 research and innovation programme under the European Research Council (ERC) Grant Agreement No. 946629, Grants No. PID2021-129035NB-I00 and PID2021- 126273NB-I00 funded by MCIN/AEI/10.13039/501100011033 and by “ERDF A way of making Europe”, and from the Basque Government through Grant No. IT1470-22.

\appendix

\section{No net momentum transfer}\label{app:no_net_momentum_tranfer}

The ``Floquet condition" Eq.~\eqref{eq:periodic_lambda_identity} and thus the time periodicity requisite of the gauge potentials imposes that both the electric and magnetic fields in Eq.~\eqref{eq:eb_field}, be time periodic without a d.c. component. This means that in order for the formalism to hold, the physically applied electromagnetic field must obey the ``no net momentum transfer" condition. As a consequence, the most general form of these fields is given by
\begin{equation}\label{eq:func_form_of_e}
\hat{\bm{E}}(\hat{\bm{r}}, t) = \sum_{n=1}^{\infty}\sum_{i=1}^d \hat{E}_n^i(\hat{\bm{r}})\cos\left(n\omega t - \varphi_n^i\right)\;\bm{e}_i,
\end{equation}
and
\begin{equation}\label{eq:func_form_of_b}
\hat{\bm{B}}(\hat{\bm{r}}, t) = \sum_{n=1}^{\infty}\sum_{i=1}^d \hat{B}_n^i(\hat{\bm{r}})\cos\left(n\omega t - \varphi_n^i\right)\;\bm{e}_i,
\end{equation}
respectively, where $d$ is the dimensionality of the system, $\bm{e}_i$ are the unitary vectors in each dimension $i$ and $\hat{E}_n^i$($\hat{B}_n^i$), $\varphi_n^i$ are the amplitude and phase corresponding to the $n$-th harmonic and the $i$-th Cartesian component, respectively.

In the case of the velocity gauge Eq.~\eqref{eq:vel_g_hamil}, $\bm{A}(t)$ is related to $\bm{E}(t)$ as
\begin{equation}
\bm{A} = -\int_{t'}^{t}\bm{E}(\tau)d\tau.
\end{equation}
Here, $t'$ is completely arbitrary and we fix it such that $\bm{A}(t') = \bm{0}$. Then, $\bm{A}$ in the velocity gauge is related to the functional form of $\bm{E}$ given in Eq.~\eqref{eq:func_form_of_e} as
\begin{equation}\label{eq:func_form_of_A}
\begin{split}
& \bm{A}(t) = \\
& \sum_{n=1}^{\infty}\sum_{i=1}^d \frac{-E_n^i}{n\omega}\left[\sin\left(n\omega t - \varphi_n^i\right)-\sin\left(n\omega t' - \varphi_n^i\right)\right]\;\bm{e}_i.
\end{split}
\end{equation}

\section{Expansion of the interaction Hamiltonian in the particle in a box example}\label{app:piab_extra}

In this appendix we consider the expansion of the velocity gauge interaction Hamiltonian of Eq.~\eqref{eq:vel_g_in_ext_systems} up to $M=2$ for the case of the particle in a box studied in Sec.~\ref{sec:particle_in_a_box}. Given that the system does not have $\bm{k}$ dependence, the resulting Hamiltonian matrix elements can be written as
\begin{equation}\label{eq:vel_2_terms}
\begin{split}
H^{V_{3}}_{nm}(t) & = E_n\delta_{nm} - \frac{q}{m}\bm{A}(t)\cdot\bm{p}_{nm} \\
& + \sum_{ij}^{d}\frac{q^2A^i(t)A^j(t)}{2m\hbar}(-i)\left[\hat{r}_j, \hat{p}_i\right]_{nm}.
\end{split}
\end{equation}
Note that since the operators $\hat{\bm{r}}$ and $\hat{\bm{p}}$ are resolved in a finite basis, $[\hat{r}_{i},\hat{p}_{j}]=i\hbar\delta_{ij}\hat{\mathds{1}}$ does not hold. This Hamiltonian can be regarded as an improved version of $H^{V_{2}}_{nm}(t)$ of Eq.~\eqref{eq:vel_m_hamil_appr}.
\begin{figure}[!tp]
\centering
\includegraphics[width=\columnwidth]{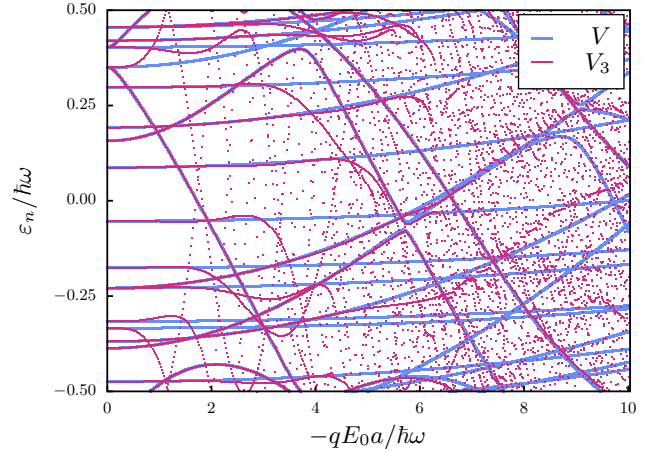}
\caption{Quasienergy spectrum of the first 20 states in the first quasienergy Brillouin zone of a system composed by a particle of charge $q$ and mass $m$ in an infinite potential well of width $a$ subjected to a driving given by the electrical field $E(t) = E_0\cos(\omega t)$. The frequency $\omega$ is fixed such that $\hbar\omega = 0.95\times 3\hbar^2\pi^2/(2ma^2)$ and $E_0$ is varied in the $[0, -10\hbar\omega/qa]$ range. The blue line shows the spectrum calculated using the velocity gauge matrix elements in the form of Eq.~\eqref{eq:vel_m_hamil} and the red line shows the spectrum calculated using Eq.~\eqref{eq:vel_2_terms}, $\varepsilon_n^V$ and $\varepsilon_n^{V_3}$, respectively.}
\label{fig:piab_vel_vs_2terms}
\end{figure}

Figure~\ref{fig:piab_vel_vs_2terms} compares the quasienergy spectrum calculated using the approximation of Eq.~\eqref{eq:vel_2_terms} with the exact result of Eq.~\eqref{eq:vel_m_hamil} (corresponding to $V$ curve in Fig.~\ref{fig:one_D_quasienergy_spectrum}). The quasienergies of the two lowest energy states that are close to resonance are perfectly described in the whole range. Other low-lying states, such as the two that display an avoided crossing at $-qE_{0}a/\hbar\omega\sim 6 $, are also very well described. In contrast, the highest energy states whose quasienergies show barely no dispersion are only well described at low values of $-qE_{0}a/\hbar\omega$. As mentioned in the main text, results can be improved in two ways; by considering additional terms in the series expansion of Eq.~\eqref{eq:vel_g_in_ext_systems} or employing a larger basis set of states.

\section{Numerical discretization of the time evolution operator}\label{app:numerical}

In this appendix we elaborate on the numerical scheme used to calculate the eigenvalues of the effective Floquet Hamiltonian. To this end, we describe the method employed to compute the one period time evolution operator $\hat{U}(t_{\text{s}} + T, t_{\text{s}})$, which then allows to obtain $\hat{H}_F(t_{\text{s}})$ from Eq.~\eqref{eq:eff_floq_hamil}. The numerical scheme discretizes the time interval $[t_{\text{s}}, t_{\text{s}} + T]$ according to
\begin{equation}
t \longrightarrow t_j = t_{\text{s}} + T\frac{j-1}{N_t-1} = t_{\text{s}} + \delta t (j-1),\quad j\in[1, N_t],
\end{equation}
and we then consider the Suzuki-Trotter expansion of the one period time evolution operator~\cite{suzukiDecompositionFormulasExponential1985},
\begin{equation}\label{eq:numerical_one_per_tev}
\begin{split}
\hat{U}(t_{\text{s}} + T, t_{\text{s}}) = & \text{exp}\left[-\frac{i}{\hbar}\sum_{j=1}^{N_t} \delta t \hat{H}(t_j) + \mathcal{O}(\delta t^2)\right]\\
= & \prod_{j=1}^{N_t}\text{exp}\left[-\frac{i}{\hbar} \delta t \hat{H}(t_j) \right]+ \mathcal{O}(\delta t^2),
\end{split}
\end{equation}
for a number of discretization points $N$.

The aim is to calculate the matrix $U(t_{\text{s}} + T, t_{\text{s}})$ of the time evolution operator, so this approach requires a matrix representation of $\hat{H}(t_j)$ for each $j$, which is considered in the main text in the context of the dipole approximation for the length and velocity gauges.

\begin{widetext}

\section{Resolution of \texorpdfstring{$\hat{\bm{r}}$}{r} in extended systems}\label{app:resolution_of_r}

In this appendix we show that the resolution of $\hat{\bm{r}}$ in the Bloch state basis leads to the definition of the covariant derivative $\hat{\bm{\mathcal{D}}}$ given in Eq.~\eqref{eq:covder} in the main text. We use the resolution of the identity Eq.~\eqref{eq:extended_resolution_of_identity} and the resolution of $\hat{\bm{r}}$ in Eq.~\eqref{eq:res_of_r_in_r_basis}, so
\begin{align}
&\hat{\bm{r}} = \int_{\text{BZ}}\int_{\text{BZ}}\frac{d^dk\;d^dq}{(2\pi)^d(2\pi)^d}\sum_{n,m=1}^{\infty}\ket{\psi_m(\bm{q})}\bra{\psi_m(\bm{q})}\hat{\bm{r}}\ket{\psi_n(\bm{k})}\bra{\psi_n(\bm{k})} = \\
&\int_{\text{BZ}}\int_{\text{BZ}}\frac{d^dk\;d^dq}{(2\pi)^d(2\pi)^d}\sum_{n,m=1}^{\infty}\int_{\mathbb{R}^d}d^d r\ket{\psi_m(\bm{q})}\braket{\psi_m(\bm{q})|\bm{r}}\bm{r}\braket{\bm{r}|\psi_n(\bm{k})}\bra{\psi_n(\bm{k})} = \\
& \int_{\text{BZ}}\int_{\text{BZ}}\frac{d^dk\;d^dq}{(2\pi)^d(2\pi)^d}\sum_{n,m=1}^{\infty}\int_{\mathbb{R}^d}d^d r\left[\bm{r}\psi_{n\bm{k}}(\bm{r})\psi_{m\bm{q}}^*(\bm{r})\right]\ket{\psi_m(\bm{q})}\bra{\psi_n(\bm{k})}\label{eq:covder_dev},
\end{align}
where in the last line we have used the definition of $\braket{\bm{r}|\psi_n(\bm{k})}$ given in Eq.~\eqref{eq:def_bloch_states}. By using the property,
\begin{equation}
\bm{\nabla}_{\bm{k}}\psi_{n\bm{k}}(\bm{r}) = i\bm{r}e^{i\bm{k}\cdot\bm{r}}u_{n\bm{k}}(\bm{r}) + e^{i\bm{k}\cdot\bm{r}}\bm{\nabla}_{\bm{k}}u_{n\bm{k}}(\bm{r})
\end{equation}
we manipulate the term inside square brackets and complete the derivative,
\begin{equation}
\begin{split}
\bm{r}\psi_{n\bm{k}}(\bm{r})\psi_{m\bm{q}}^*(\bm{r}) = & \left(-i\bm{\nabla}_{\bm{k}}\psi_{n\bm{k}}(\bm{r}) + ie^{i\bm{k}\cdot\bm{r}}\bm{\nabla}_{\bm{k}}u_{n\bm{k}}(\bm{r})\right)\psi_{m\bm{q}}^*(\bm{r}) = \\
&(-i)\bm{\nabla}_{\bm{k}}\psi_{m\bm{q}}^*(\bm{r})\psi_{n\bm{k}}(\bm{r}) + i e^{i(\bm{k}-\bm{q})\cdot\bm{r}}u_{m\bm{q}}^*(\bm{r})\bm{\nabla}_{\bm{k}}u_{n\bm{k}}(\bm{r}).
\end{split}
\end{equation}
We now integrate over $\bm{r}$ each term in this sum separately,
\begin{align}
&\int_{\mathbb{R}^d}d^d r (-i)\bm{\nabla}_{\bm{k}}\psi_{m\bm{q}}^*(\bm{r})\psi_{n\bm{k}}(\bm{r}) = (-i)\bm{\nabla}_{\bm{k}}\int_{\mathbb{R}^d}d^d r \psi_{m\bm{q}}^*(\bm{r})\psi_{n\bm{k}}(\bm{r}) =\label{eq:covder_der_term_dev_1}\\
&(-i)\bm{\nabla}_{\bm{k}} \braket{\psi_m(\bm{q})|\psi_n(\bm{k})} = (-i)(2\pi)^d\delta_{mn}\bm{\nabla}_{\bm{k}}\delta(\bm{k}-\bm{q})\label{eq:covder_der_term_dev_2},
\end{align}
and
\begin{equation}
i\int_{\mathbb{R}^d}d^d r e^{i(\bm{k}-\bm{q})\cdot\bm{r}}u_{m\bm{q}}(\bm{r})\bm{\nabla}_{\bm{k}}u_{n\bm{k}}(\bm{r}) = (2\pi)^d \delta(\bm{k}-\bm{q})\frac{i}{\Omega_{\text{cell}}}\int_{\Omega_{\text{cell}}}d^d r u_{m\bm{q}}^*(\bm{r})\bm{\nabla}_{\bm{k}}u_{n\bm{k}}(\bm{r}) = (2\pi)^d \delta(\bm{k}-\bm{q})\bm{\xi}(\bm{k}),
\end{equation}
where $\Omega_{\text{cell}}$ is the unit cell volume, and $\bm{\xi}$ is the Berry connection, also defined in Eq.~\eqref{eq:berry_conn} in the main text. So,
\begin{equation}
\int_{\mathbb{R}^d}d^d r\bm{r}\psi_{n\bm{k}}(\bm{r})\psi_{m\bm{q}}^*(\bm{r}) = (-i)(2\pi)^d\delta_{mn}\bm{\nabla}_{\bm{k}}\delta(\bm{k}-\bm{q}) + (2\pi)^d \delta(\bm{k}-\bm{q})\bm{\xi}(\bm{k}).
\end{equation}
Plugging this result into Eq.~\eqref{eq:covder_dev} leads to the definition of $\hat{\bm{\mathcal{D}}}$ given in Eq.~\eqref{eq:covder} in the main text.

\end{widetext}

\section{Wannier interpolation}\label{app:wannier}

At the implementation level, we have resolved the operators mentioned in the main text in the quantum-geometric gauge known as the Wannier gauge. The idea is that localized Wannier functions $\ket{n\bm{R}}$ can be used to construct Bloch states~\cite{vanderbiltBerryPhasesElectronic2018}
\begin{equation}\label{eq:wannier_gauge}
\ket{\psi_n^{\text{W}}(\bm{k})} = \sum_{\bm{R}}e^{i\bm{k}\cdot\bm{R}}\ket{n\bm{R}},
\end{equation}
which provide a basis to resolve $\hat{H}_0$ and $\hat{\xi}^j$. In practice, the Wannier functions $\ket{n\bm{R}}$ are calculated using the \textsc{Wannier90}~\cite{pizziWannier90CommunityCode2020} package. In general, the $\ket{\psi_n^{\text{W}}(\bm{k})}$'s are not eigenstates of any operator, but provide a smooth basis with respect to $\bm{k}$ that allows for interpolation of k-dependent quantities, the so-called Wannier interpolation. The basic building blocks of the formalism are the position and Hamiltonian matrix elements in Wannier basis $H^{\text{W}}_{0\;nm}(\bm{R}) = \braket{n\bm{0}|\hat{H}_0|m\bm{R}}$ and $\xi^{j\;\text{W}}(\bm{R})= \braket{n\bm{0}|\hat{\xi}|m\bm{R}}$, respectively. Their Fourier transforms yield
\begin{align}
H^{\text{W}}_{0}(\bm{k}) = &\sum_{\bm{R}}e^{i\bm{k}\cdot\bm{R}}H^{\text{W}}_{0}(\bm{R}),\\
\xi^{j\;\text{W}}(\bm{k}) = &\sum_{\bm{R}}e^{i\bm{k}\cdot\bm{R}}\xi^{j\;\text{W}}(\bm{R}),
\end{align}
and k-space derivatives of these terms can be computed straightforwardly~\cite{wangInitioCalculationAnomalous2006,yatesSpectralFermiSurface2007,ibanez-azpirozInitioCalculationShift2018}.

Once these quantities are computed in a certain quantum-geometric gauge, we have means to resolve $\hat{H}(t)$. In the velocity gauge expression, each individual term of the expansion Eq.~\eqref{eq:vel_g_in_ext_systems} is explicitly quantum-geometric gauge-covariant. Therefore, we have proceeded by using the Wannier interpolation technique to compute the matrices representing $\hat{\mathcal{D}}^{\alpha_1}\cdots\hat{\mathcal{D}}^{\alpha_n}[\hat{H}_0]$ in the Wannier gauge. Then, we obtain the matrices of $H^{\text{W}}(\bm{k}, t)$, $U^{\text{W}}(\bm{k}, t_{\text{s}}+T, t_{\text{s}})$, and $H_{\text{F}}^{\text{W}}(\bm{k}, t_{\text{s}})$ in the Wannier gauge by using Eqs.~\eqref{eq:vel_g_in_ext_systems},~\eqref{eq:numerical_one_per_tev} and ~\eqref{eq:eff_floq_hamil}, respectively. For the computation of the expansion terms of Eq.~\eqref{eq:vel_g_in_ext_systems}, we have used
\begin{equation}
(\hat{\mathcal{D}}^{j}[\hat{H}_0])^{\text{W}}_{nm} = \frac{\partial H^{\text{W}}_{0\;nm}}{\partial k_j} - i[\xi^{j\;\text{W}}, H^{\text{W}}_{0}]_{nm},
\end{equation}
and the analogous expressions for higher order terms $(\hat{\mathcal{D}}^{l}\hat{\mathcal{D}}^{j}[\hat{H}_0])^{\text{W}}_{nm},\;\cdots$ by employing Eq.~\eqref{eq:covder_action_on_operator}. In the case of the length gauge approximation of Eq.~\eqref{eq:r-app}, $\hat{H}(t)$ is not gauge-covariant and is evaluated in the quantum-geometric gauge that diagonalizes the Hamiltonian. In this case, we first calculate the matrices $H_0^{\text{W}}(\bm{k})$ and $\xi^{j\;\text{W}}(\bm{k})$ in the Wannier gauge and transform them to a new gauge by diagonalizing $H_0^{\text{W}}(\bm{k})$ and using their respective gauge transformation properties. That is, for this case we define the gauge transformation $\mathcal{U}(\bm{k})$ in Eq.~\eqref{eq:quantum_geometric_gauge_freedom} as the gauge transformation diagonalizing $H_0^{\text{W}}(\bm{k})$. This defines the so called ``Hamiltonian" gauge H, where the matrices $H_0^{\text{H}}(\bm{k})$ and $\xi^{j\;\text{H}}(\bm{k})$ are calculated. It is in this gauge that we compute the approximation Eq.~\eqref{eq:r-app} and calculate, using the length gauge, the matrix $H^{\text{H}}(\bm{k}, t)$ in the quantum-geometric Hamiltonian gauge. We then proceed as in the velocity gauge case and calculate the matrix $H^{\text{H}}_{\text{F}}(\bm{k})$, which as the superscript indicates is calculated in the Hamiltonian gauge.

\section{Time complexity analysis}\label{app:time_complexity}
In this appendix we briefly analyze the time complexity related to the calculation of the quasienergy spectrum. We start by analyzing the complexity of the method in terms of the internally fixed variables, namely those variables which do not correspond to the physical setting of the calculation and just control the precision. These include the number of states of the crystal spectrum, $N$, leading to $N\times N$ matrices representing $\hat{H}_0$ and $\hat{\xi}^j$. The number of steps for the discretization of the one-cycle time evolution operator in Eq.~\eqref{eq:numerical_one_per_tev}, $N_t$, and the calculation method employed to resolve $\hat{H}(t_j)$, where we distinguish between finite and extended systems. In the case of extended systems we employ Eq.~\eqref{eq:vel_g_in_ext_systems} (ExV) and in the case of finite systems we further distinguish between velocity and length gauge expressions, Eqs.~\eqref{eq:vel_g_hamil} (FnV) and~\eqref{eq:len_g_hamil} (FnE), respectively.

By virtue of Eq.~\eqref{eq:numerical_one_per_tev}, the computation of the one-cycle time evolution operator requires computing $N_t$ matrix exponentials of size $N\times N$, which are then multiplied sequentially. Then, by virtue of Eq.~\eqref{eq:eff_floq_hamil}, a further diagonalization is required to obtain the quasienergy spectrum. Each $N\times N$ matrix multiplication has, at most, a time complexity of $\mathcal{O}(N^3)$~\cite{highamAccuracyStabilityNumerical2002}, while the diagonalization procedure, carried out by \textsc{Lapack}'s \textsc{Zheev} routine, which employs the QR algorithm, involves a time complexity of $\mathcal{O}(N^3)$~\cite{highamAccuracyStabilityNumerical2002}. The matrix exponential operation involves two matrix multiplications and a single diagonalization. This leads to a complexity
\begin{equation}
\begin{split}
& \mathcal{O}(C)+\mathcal{O}(N_t (N^3 + 2N^3 + N^3) + N^3) = \\
& \mathcal{O}(C)+\mathcal{O}((4N_t + 1)N^3),
\end{split}
\end{equation}
where $C$ is the time complexity associated with the resolution of $\hat{H}(t_j)$ in any of the FnE, FnV and ExV cases.

The FnE case requires adding $d+1$ matrices, thus
\begin{equation}
C^{\text{FnE}} = \mathcal{O}(dN^2),
\end{equation}
the FnV case requires computing two matrix exponentials and two matrix multiplications, leading to
\begin{equation}
C^{\text{FnV}} = \mathcal{O}(2\times3N^3 + 2N^3) = \mathcal{O}(8N^3).
\end{equation}
Lastly, the ExV case requires computing $M$ objects, the successive actions $\hat{D}^{\alpha_M}\cdots\hat{D}^{\alpha_1}[\hat{H}_0]$, each of which can be considered an $N\times N\times d^l$ array, where $l\in[1, M]$. Each object is contracted $d\times l$ times with the $A_{\alpha_j}(t)$s, the vector potential components, obtaining $M$ matrices of size $N\times N$, which are then summed together. The calculation of a single $\hat{D}^{\alpha_l}\cdots\hat{D}^{\alpha_1}[\hat{H}_0]$ array involves summing $l$-times nested commutators, thus it involves a computational complexity of $\mathcal{O}(l\times N^3)$. The contraction with the $A_{\alpha_j}(t)$s involves a computational complexity of $\mathcal{O}(d\times l\times N^2)$ and the addition of $M$ matrices involves a computational complexity of $\mathcal{O}(M\times N^2)$. Thus, ignoring $\mathcal{O}(N^2)$ contributions,
\begin{equation}
\begin{split}
& C^{\text{ExV}} = \mathcal{O}\left(N^3\times\sum_{l=1}^M l\right) = \\
& \mathcal{O}\left(\frac{(M+1)M}{2}N^3\right).
\end{split}
\end{equation}

Finally, we deal with the functional dependence that the quasienergy spectrum has with respect to externally fixed variables, namely with respect to the amplitudes $E_n^i$, phases $\varphi_n^i$ and frequency $\omega$ in Eq.~\eqref{eq:func_form_of_e}. This accounts for $2\times N_h\times d + 1$ variables, where $N_h$ stands for the number of harmonics considered in Eq.~\eqref{eq:func_form_of_e} and $d$ for the dimensionality. Finally, the spectrum also depends functionally on $\bm{k}$ in the case of extended systems, thus
\begin{equation}
\varepsilon_n \equiv \varepsilon_n\left(E_1^1, \cdots, E_{N_h}^d, \varphi_1^1, \cdots, \varphi_{N_h}^d, \omega, \bm{k}\right).
\end{equation}
Each variation of any of the external variables involves a multiplicative increase of time complexity, since the algorithm must be run again for each external variable permutation.

\bibliography{bibdata}

\begin{thebibliography}{64}%
\makeatletter
\providecommand \@ifxundefined [1]{%
 \@ifx{#1\undefined}
}%
\providecommand \@ifnum [1]{%
 \ifnum #1\expandafter \@firstoftwo
 \else \expandafter \@secondoftwo
 \fi
}%
\providecommand \@ifx [1]{%
 \ifx #1\expandafter \@firstoftwo
 \else \expandafter \@secondoftwo
 \fi
}%
\providecommand \natexlab [1]{#1}%
\providecommand \enquote  [1]{``#1''}%
\providecommand \bibnamefont  [1]{#1}%
\providecommand \bibfnamefont [1]{#1}%
\providecommand \citenamefont [1]{#1}%
\providecommand \href@noop [0]{\@secondoftwo}%
\providecommand \href [0]{\begingroup \@sanitize@url \@href}%
\providecommand \@href[1]{\@@startlink{#1}\@@href}%
\providecommand \@@href[1]{\endgroup#1\@@endlink}%
\providecommand \@sanitize@url [0]{\catcode `\\12\catcode `\$12\catcode
  `\&12\catcode `\#12\catcode `\^12\catcode `\_12\catcode `\%12\relax}%
\providecommand \@@startlink[1]{}%
\providecommand \@@endlink[0]{}%
\providecommand \url  [0]{\begingroup\@sanitize@url \@url }%
\providecommand \@url [1]{\endgroup\@href {#1}{\urlprefix }}%
\providecommand \urlprefix  [0]{URL }%
\providecommand \Eprint [0]{\href }%
\providecommand \doibase [0]{https://doi.org/}%
\providecommand \selectlanguage [0]{\@gobble}%
\providecommand \bibinfo  [0]{\@secondoftwo}%
\providecommand \bibfield  [0]{\@secondoftwo}%
\providecommand \translation [1]{[#1]}%
\providecommand \BibitemOpen [0]{}%
\providecommand \bibitemStop [0]{}%
\providecommand \bibitemNoStop [0]{.\EOS\space}%
\providecommand \EOS [0]{\spacefactor3000\relax}%
\providecommand \BibitemShut  [1]{\csname bibitem#1\endcsname}%
\let\auto@bib@innerbib\@empty
\bibitem [{\citenamefont {Sambe}(1973)}]{sambeSteadyStatesQuasienergies1973}%
  \BibitemOpen
  \bibfield  {author} {\bibinfo {author} {\bibfnamefont {H.}~\bibnamefont
  {Sambe}},\ }\bibfield  {title} {\bibinfo {title} {Steady {{States}} and
  {{Quasienergies}} of a {{Quantum-Mechanical System}} in an {{Oscillating
  Field}}},\ }\href {https://doi.org/10.1103/PhysRevA.7.2203} {\bibfield
  {journal} {\bibinfo  {journal} {Physical Review A}\ }\textbf {\bibinfo
  {volume} {7}},\ \bibinfo {pages} {2203} (\bibinfo {year} {1973})}\BibitemShut
  {NoStop}%
\bibitem [{\citenamefont {Chu}(2007)}]{chuGeneralizedFloquetTheoretical2007}%
  \BibitemOpen
  \bibfield  {author} {\bibinfo {author} {\bibfnamefont {S.-I.}\ \bibnamefont
  {Chu}},\ }\bibfield  {title} {\bibinfo {title} {Generalized {{Floquet
  Theoretical Approaches}} to {{Intense-Field Multiphoton}} and {{Nonlinear
  Optical Processes}}},\ }in\ \href
  {https://doi.org/10.1002/9780470141229.ch17} {\emph {\bibinfo {booktitle}
  {Advances in {{Chemical Physics}}}}},\ \bibinfo {editor} {edited by\ \bibinfo
  {editor} {\bibfnamefont {J.~O.}\ \bibnamefont {Hirschfelder}}, \bibinfo
  {editor} {\bibfnamefont {R.~E.}\ \bibnamefont {Wyatt}},\ and\ \bibinfo
  {editor} {\bibfnamefont {R.~D.}\ \bibnamefont {Coalson}}}\ (\bibinfo
  {publisher} {John Wiley \& Sons, Inc.},\ \bibinfo {address} {Hoboken, NJ,
  USA},\ \bibinfo {year} {2007})\ pp.\ \bibinfo {pages} {739--799}\BibitemShut
  {NoStop}%
\bibitem [{\citenamefont {Fainshtein}\ \emph {et~al.}(1978)\citenamefont
  {Fainshtein}, \citenamefont {Manakov},\ and\ \citenamefont
  {Rapoport}}]{fainshteinGeneralPropertiesQuasienergetic1978}%
  \BibitemOpen
  \bibfield  {author} {\bibinfo {author} {\bibfnamefont {A.~G.}\ \bibnamefont
  {Fainshtein}}, \bibinfo {author} {\bibfnamefont {N.~L.}\ \bibnamefont
  {Manakov}},\ and\ \bibinfo {author} {\bibfnamefont {L.~P.}\ \bibnamefont
  {Rapoport}},\ }\bibfield  {title} {\bibinfo {title} {Some general properties
  of quasi-energetic spectra of quantum systems in classical monochromatic
  fields},\ }\href {https://doi.org/10.1088/0022-3700/11/14/020} {\bibfield
  {journal} {\bibinfo  {journal} {Journal of Physics B: Atomic and Molecular
  Physics}\ }\textbf {\bibinfo {volume} {11}},\ \bibinfo {pages} {2561}
  (\bibinfo {year} {1978})}\BibitemShut {NoStop}%
\bibitem [{\citenamefont {Weinberg}\ \emph {et~al.}(2015)\citenamefont
  {Weinberg}, \citenamefont {{\"O}lschl{\"a}ger}, \citenamefont {Str{\"a}ter},
  \citenamefont {Prelle}, \citenamefont {Eckardt}, \citenamefont {Sengstock},\
  and\ \citenamefont {Simonet}}]{weinbergMultiphotonInterbandExcitations2015}%
  \BibitemOpen
  \bibfield  {author} {\bibinfo {author} {\bibfnamefont {M.}~\bibnamefont
  {Weinberg}}, \bibinfo {author} {\bibfnamefont {C.}~\bibnamefont
  {{\"O}lschl{\"a}ger}}, \bibinfo {author} {\bibfnamefont {C.}~\bibnamefont
  {Str{\"a}ter}}, \bibinfo {author} {\bibfnamefont {S.}~\bibnamefont {Prelle}},
  \bibinfo {author} {\bibfnamefont {A.}~\bibnamefont {Eckardt}}, \bibinfo
  {author} {\bibfnamefont {K.}~\bibnamefont {Sengstock}},\ and\ \bibinfo
  {author} {\bibfnamefont {J.}~\bibnamefont {Simonet}},\ }\bibfield  {title}
  {\bibinfo {title} {Multiphoton interband excitations of quantum gases in
  driven optical lattices},\ }\href
  {https://doi.org/10.1103/PhysRevA.92.043621} {\bibfield  {journal} {\bibinfo
  {journal} {Physical Review A}\ }\textbf {\bibinfo {volume} {92}},\ \bibinfo
  {pages} {043621} (\bibinfo {year} {2015})}\BibitemShut {NoStop}%
\bibitem [{\citenamefont {Dunlap}\ and\ \citenamefont
  {Kenkre}(1986)}]{dunlapDynamicLocalizationCharged1986}%
  \BibitemOpen
  \bibfield  {author} {\bibinfo {author} {\bibfnamefont {D.~H.}\ \bibnamefont
  {Dunlap}}\ and\ \bibinfo {author} {\bibfnamefont {V.~M.}\ \bibnamefont
  {Kenkre}},\ }\bibfield  {title} {\bibinfo {title} {Dynamic localization of a
  charged particle moving under the influence of an electric field},\ }\href
  {https://doi.org/10.1103/PhysRevB.34.3625} {\bibfield  {journal} {\bibinfo
  {journal} {Physical Review B}\ }\textbf {\bibinfo {volume} {34}},\ \bibinfo
  {pages} {3625} (\bibinfo {year} {1986})}\BibitemShut {NoStop}%
\bibitem [{\citenamefont {Keshavamurthy}\ and\ \citenamefont
  {Schlagheck}(2011)}]{keshavamurthyDynamicalTunnelingTheory2011}%
  \BibitemOpen
  \bibinfo {editor} {\bibfnamefont {S.}~\bibnamefont {Keshavamurthy}}\ and\
  \bibinfo {editor} {\bibfnamefont {P.}~\bibnamefont {Schlagheck}},\ eds.,\
  \href {https://doi.org/10.1201/b10712} {\emph {\bibinfo {title} {Dynamical
  {{Tunneling}}: {{Theory}} and {{Experiment}}}}},\ \bibinfo {edition} {1st}\
  ed.\ (\bibinfo  {publisher} {CRC Press},\ \bibinfo {address} {Boca Raton},\
  \bibinfo {year} {2011})\BibitemShut {NoStop}%
\bibitem [{\citenamefont {Arlinghaus}\ \emph {et~al.}(2011)\citenamefont
  {Arlinghaus}, \citenamefont {Langemeyer},\ and\ \citenamefont
  {Holthaus}}]{arlinghausDynamicLocalizationOptical2011}%
  \BibitemOpen
  \bibfield  {author} {\bibinfo {author} {\bibfnamefont {S.}~\bibnamefont
  {Arlinghaus}}, \bibinfo {author} {\bibfnamefont {M.}~\bibnamefont
  {Langemeyer}},\ and\ \bibinfo {author} {\bibfnamefont {M.}~\bibnamefont
  {Holthaus}},\ }\bibfield  {title} {\bibinfo {title} {Dynamic {{Localization}}
  in {{Optical Lattices}}},\ }\bibfield  {journal} {\bibinfo  {journal}
  {arXiv}\ }\href {https://doi.org/10.48550/ARXIV.1103.4293}
  {10.48550/ARXIV.1103.4293} (\bibinfo {year} {2011})\BibitemShut {NoStop}%
\bibitem [{\citenamefont {Eckardt}(2017)}]{eckardtColloquiumAtomicQuantum2017}%
  \BibitemOpen
  \bibfield  {author} {\bibinfo {author} {\bibfnamefont {A.}~\bibnamefont
  {Eckardt}},\ }\bibfield  {title} {\bibinfo {title} {Colloquium: {{Atomic}}
  quantum gases in periodically driven optical lattices},\ }\href
  {https://doi.org/10.1103/RevModPhys.89.011004} {\bibfield  {journal}
  {\bibinfo  {journal} {Reviews of Modern Physics}\ }\textbf {\bibinfo {volume}
  {89}},\ \bibinfo {pages} {011004} (\bibinfo {year} {2017})}\BibitemShut
  {NoStop}%
\bibitem [{\citenamefont {Kitagawa}\ \emph {et~al.}(2010)\citenamefont
  {Kitagawa}, \citenamefont {Berg}, \citenamefont {Rudner},\ and\ \citenamefont
  {Demler}}]{kitagawaTopologicalCharacterizationPeriodically2010}%
  \BibitemOpen
  \bibfield  {author} {\bibinfo {author} {\bibfnamefont {T.}~\bibnamefont
  {Kitagawa}}, \bibinfo {author} {\bibfnamefont {E.}~\bibnamefont {Berg}},
  \bibinfo {author} {\bibfnamefont {M.}~\bibnamefont {Rudner}},\ and\ \bibinfo
  {author} {\bibfnamefont {E.}~\bibnamefont {Demler}},\ }\bibfield  {title}
  {\bibinfo {title} {Topological characterization of periodically driven
  quantum systems},\ }\href {https://doi.org/10.1103/PhysRevB.82.235114}
  {\bibfield  {journal} {\bibinfo  {journal} {Physical Review B}\ }\textbf
  {\bibinfo {volume} {82}},\ \bibinfo {pages} {235114} (\bibinfo {year}
  {2010})}\BibitemShut {NoStop}%
\bibitem [{\citenamefont {Nakagawa}\ \emph {et~al.}(2020)\citenamefont
  {Nakagawa}, \citenamefont {Slager}, \citenamefont {Higashikawa},\ and\
  \citenamefont {Oka}}]{nakagawaWannierRepresentationFloquet2020}%
  \BibitemOpen
  \bibfield  {author} {\bibinfo {author} {\bibfnamefont {M.}~\bibnamefont
  {Nakagawa}}, \bibinfo {author} {\bibfnamefont {R.-J.}\ \bibnamefont
  {Slager}}, \bibinfo {author} {\bibfnamefont {S.}~\bibnamefont
  {Higashikawa}},\ and\ \bibinfo {author} {\bibfnamefont {T.}~\bibnamefont
  {Oka}},\ }\bibfield  {title} {\bibinfo {title} {Wannier representation of
  {{Floquet}} topological states},\ }\href
  {https://doi.org/10.1103/PhysRevB.101.075108} {\bibfield  {journal} {\bibinfo
   {journal} {Physical Review B}\ }\textbf {\bibinfo {volume} {101}},\ \bibinfo
  {pages} {075108} (\bibinfo {year} {2020})}\BibitemShut {NoStop}%
\bibitem [{\citenamefont {{Puente-Uriona}}\ \emph {et~al.}(2024)\citenamefont
  {{Puente-Uriona}}, \citenamefont {Pettini},\ and\ \citenamefont
  {Modugno}}]{puente-urionaTopologicalPhaseDiagram2024}%
  \BibitemOpen
  \bibfield  {author} {\bibinfo {author} {\bibfnamefont {{\'A}.~R.}\
  \bibnamefont {{Puente-Uriona}}}, \bibinfo {author} {\bibfnamefont
  {G.}~\bibnamefont {Pettini}},\ and\ \bibinfo {author} {\bibfnamefont
  {M.}~\bibnamefont {Modugno}},\ }\bibfield  {title} {\bibinfo {title}
  {Topological phase diagram of optimally shaken honeycomb lattices: {{A}} dual
  perspective from stroboscopic and nonstroboscopic {{Floquet Hamiltonians}}},\
  }\href {https://doi.org/10.1103/PhysRevResearch.6.023244} {\bibfield
  {journal} {\bibinfo  {journal} {Physical Review Research}\ }\textbf {\bibinfo
  {volume} {6}},\ \bibinfo {pages} {023244} (\bibinfo {year}
  {2024})}\BibitemShut {NoStop}%
\bibitem [{\citenamefont {Grushin}\ \emph {et~al.}(2014)\citenamefont
  {Grushin}, \citenamefont {{G{\'o}mez-Le{\'o}n}},\ and\ \citenamefont
  {Neupert}}]{grushinFloquetFractionalChern2014}%
  \BibitemOpen
  \bibfield  {author} {\bibinfo {author} {\bibfnamefont {A.~G.}\ \bibnamefont
  {Grushin}}, \bibinfo {author} {\bibfnamefont {{\'A}.}~\bibnamefont
  {{G{\'o}mez-Le{\'o}n}}},\ and\ \bibinfo {author} {\bibfnamefont
  {T.}~\bibnamefont {Neupert}},\ }\bibfield  {title} {\bibinfo {title} {Floquet
  {{Fractional Chern Insulators}}},\ }\href
  {https://doi.org/10.1103/PhysRevLett.112.156801} {\bibfield  {journal}
  {\bibinfo  {journal} {Physical Review Letters}\ }\textbf {\bibinfo {volume}
  {112}},\ \bibinfo {pages} {156801} (\bibinfo {year} {2014})}\BibitemShut
  {NoStop}%
\bibitem [{\citenamefont {Morimoto}\ and\ \citenamefont
  {Nagaosa}(2016)}]{morimotoTopologicalNatureNonlinear2016}%
  \BibitemOpen
  \bibfield  {author} {\bibinfo {author} {\bibfnamefont {T.}~\bibnamefont
  {Morimoto}}\ and\ \bibinfo {author} {\bibfnamefont {N.}~\bibnamefont
  {Nagaosa}},\ }\bibfield  {title} {\bibinfo {title} {Topological nature of
  nonlinear optical effects in solids},\ }\href
  {https://doi.org/10.1126/sciadv.1501524} {\bibfield  {journal} {\bibinfo
  {journal} {Science Advances}\ }\textbf {\bibinfo {volume} {2}},\ \bibinfo
  {pages} {e1501524} (\bibinfo {year} {2016})}\BibitemShut {NoStop}%
\bibitem [{\citenamefont {Goldman}\ and\ \citenamefont
  {Dalibard}(2014)}]{goldmanPeriodicallyDrivenQuantum2014}%
  \BibitemOpen
  \bibfield  {author} {\bibinfo {author} {\bibfnamefont {N.}~\bibnamefont
  {Goldman}}\ and\ \bibinfo {author} {\bibfnamefont {J.}~\bibnamefont
  {Dalibard}},\ }\bibfield  {title} {\bibinfo {title} {Periodically {{Driven
  Quantum Systems}}: {{Effective Hamiltonians}} and {{Engineered Gauge
  Fields}}},\ }\href {https://doi.org/10.1103/PhysRevX.4.031027} {\bibfield
  {journal} {\bibinfo  {journal} {Physical Review X}\ }\textbf {\bibinfo
  {volume} {4}},\ \bibinfo {pages} {031027} (\bibinfo {year}
  {2014})}\BibitemShut {NoStop}%
\bibitem [{\citenamefont {Bukov}\ \emph {et~al.}(2015)\citenamefont {Bukov},
  \citenamefont {D'Alessio},\ and\ \citenamefont
  {Polkovnikov}}]{bukovUniversalHighfrequencyBehavior2015}%
  \BibitemOpen
  \bibfield  {author} {\bibinfo {author} {\bibfnamefont {M.}~\bibnamefont
  {Bukov}}, \bibinfo {author} {\bibfnamefont {L.}~\bibnamefont {D'Alessio}},\
  and\ \bibinfo {author} {\bibfnamefont {A.}~\bibnamefont {Polkovnikov}},\
  }\bibfield  {title} {\bibinfo {title} {Universal high-frequency behavior of
  periodically driven systems: From dynamical stabilization to {{Floquet}}
  engineering},\ }\href {https://doi.org/10.1080/00018732.2015.1055918}
  {\bibfield  {journal} {\bibinfo  {journal} {Advances in Physics}\ }\textbf
  {\bibinfo {volume} {64}},\ \bibinfo {pages} {139} (\bibinfo {year}
  {2015})}\BibitemShut {NoStop}%
\bibitem [{\citenamefont {Castro}\ \emph {et~al.}(2023)\citenamefont {Castro},
  \citenamefont {Giovannini}, \citenamefont {Sato}, \citenamefont
  {H{\"u}bener},\ and\ \citenamefont
  {Rubio}}]{castroFloquetEngineeringQuantum2023}%
  \BibitemOpen
  \bibfield  {author} {\bibinfo {author} {\bibfnamefont {A.}~\bibnamefont
  {Castro}}, \bibinfo {author} {\bibfnamefont {U.~D.}\ \bibnamefont
  {Giovannini}}, \bibinfo {author} {\bibfnamefont {S.~A.}\ \bibnamefont
  {Sato}}, \bibinfo {author} {\bibfnamefont {H.}~\bibnamefont {H{\"u}bener}},\
  and\ \bibinfo {author} {\bibfnamefont {A.}~\bibnamefont {Rubio}},\ }\bibfield
   {title} {\bibinfo {title} {Floquet engineering with quantum optimal control
  theory},\ }\href {https://doi.org/10.1088/1367-2630/accb05} {\bibfield
  {journal} {\bibinfo  {journal} {New Journal of Physics}\ }\textbf {\bibinfo
  {volume} {25}},\ \bibinfo {pages} {043023} (\bibinfo {year}
  {2023})}\BibitemShut {NoStop}%
\bibitem [{\citenamefont {Jotzu}\ \emph {et~al.}(2014)\citenamefont {Jotzu},
  \citenamefont {Messer}, \citenamefont {Desbuquois}, \citenamefont {Lebrat},
  \citenamefont {Uehlinger}, \citenamefont {Greif},\ and\ \citenamefont
  {Esslinger}}]{jotzuExperimentalRealizationTopological2014}%
  \BibitemOpen
  \bibfield  {author} {\bibinfo {author} {\bibfnamefont {G.}~\bibnamefont
  {Jotzu}}, \bibinfo {author} {\bibfnamefont {M.}~\bibnamefont {Messer}},
  \bibinfo {author} {\bibfnamefont {R.}~\bibnamefont {Desbuquois}}, \bibinfo
  {author} {\bibfnamefont {M.}~\bibnamefont {Lebrat}}, \bibinfo {author}
  {\bibfnamefont {T.}~\bibnamefont {Uehlinger}}, \bibinfo {author}
  {\bibfnamefont {D.}~\bibnamefont {Greif}},\ and\ \bibinfo {author}
  {\bibfnamefont {T.}~\bibnamefont {Esslinger}},\ }\bibfield  {title} {\bibinfo
  {title} {Experimental realization of the topological {{Haldane}} model with
  ultracold fermions},\ }\href {https://doi.org/10.1038/nature13915} {\bibfield
   {journal} {\bibinfo  {journal} {Nature}\ }\textbf {\bibinfo {volume}
  {515}},\ \bibinfo {pages} {237} (\bibinfo {year} {2014})}\BibitemShut
  {NoStop}%
\bibitem [{\citenamefont {Modugno}\ and\ \citenamefont
  {Pettini}(2017)}]{modugnoCorrespondenceShakenHoneycomb2017}%
  \BibitemOpen
  \bibfield  {author} {\bibinfo {author} {\bibfnamefont {M.}~\bibnamefont
  {Modugno}}\ and\ \bibinfo {author} {\bibfnamefont {G.}~\bibnamefont
  {Pettini}},\ }\bibfield  {title} {\bibinfo {title} {Correspondence between a
  shaken honeycomb lattice and the {{Haldane}} model},\ }\href
  {https://doi.org/10.1103/PhysRevA.96.053603} {\bibfield  {journal} {\bibinfo
  {journal} {Physical Review A}\ }\textbf {\bibinfo {volume} {96}},\ \bibinfo
  {pages} {053603} (\bibinfo {year} {2017})}\BibitemShut {NoStop}%
\bibitem [{\citenamefont {Eckardt}\ \emph {et~al.}(2005)\citenamefont
  {Eckardt}, \citenamefont {Weiss},\ and\ \citenamefont
  {Holthaus}}]{eckardtSuperfluidInsulatorTransitionPeriodically2005}%
  \BibitemOpen
  \bibfield  {author} {\bibinfo {author} {\bibfnamefont {A.}~\bibnamefont
  {Eckardt}}, \bibinfo {author} {\bibfnamefont {C.}~\bibnamefont {Weiss}},\
  and\ \bibinfo {author} {\bibfnamefont {M.}~\bibnamefont {Holthaus}},\
  }\bibfield  {title} {\bibinfo {title} {Superfluid-{{Insulator Transition}} in
  a {{Periodically Driven Optical Lattice}}},\ }\href
  {https://doi.org/10.1103/PhysRevLett.95.260404} {\bibfield  {journal}
  {\bibinfo  {journal} {Physical Review Letters}\ }\textbf {\bibinfo {volume}
  {95}},\ \bibinfo {pages} {260404} (\bibinfo {year} {2005})}\BibitemShut
  {NoStop}%
\bibitem [{\citenamefont {Zenesini}\ \emph {et~al.}(2009)\citenamefont
  {Zenesini}, \citenamefont {Lignier}, \citenamefont {Ciampini}, \citenamefont
  {Morsch},\ and\ \citenamefont
  {Arimondo}}]{zenesiniCoherentControlDressed2009}%
  \BibitemOpen
  \bibfield  {author} {\bibinfo {author} {\bibfnamefont {A.}~\bibnamefont
  {Zenesini}}, \bibinfo {author} {\bibfnamefont {H.}~\bibnamefont {Lignier}},
  \bibinfo {author} {\bibfnamefont {D.}~\bibnamefont {Ciampini}}, \bibinfo
  {author} {\bibfnamefont {O.}~\bibnamefont {Morsch}},\ and\ \bibinfo {author}
  {\bibfnamefont {E.}~\bibnamefont {Arimondo}},\ }\bibfield  {title} {\bibinfo
  {title} {Coherent {{Control}} of {{Dressed Matter Waves}}},\ }\href
  {https://doi.org/10.1103/PhysRevLett.102.100403} {\bibfield  {journal}
  {\bibinfo  {journal} {Physical Review Letters}\ }\textbf {\bibinfo {volume}
  {102}},\ \bibinfo {pages} {100403} (\bibinfo {year} {2009})}\BibitemShut
  {NoStop}%
\bibitem [{\citenamefont {Jiang}\ \emph {et~al.}(2011)\citenamefont {Jiang},
  \citenamefont {Kitagawa}, \citenamefont {Alicea}, \citenamefont {Akhmerov},
  \citenamefont {Pekker}, \citenamefont {Refael}, \citenamefont {Cirac},
  \citenamefont {Demler}, \citenamefont {Lukin},\ and\ \citenamefont
  {Zoller}}]{jiangMajoranaFermionsEquilibrium2011}%
  \BibitemOpen
  \bibfield  {author} {\bibinfo {author} {\bibfnamefont {L.}~\bibnamefont
  {Jiang}}, \bibinfo {author} {\bibfnamefont {T.}~\bibnamefont {Kitagawa}},
  \bibinfo {author} {\bibfnamefont {J.}~\bibnamefont {Alicea}}, \bibinfo
  {author} {\bibfnamefont {A.~R.}\ \bibnamefont {Akhmerov}}, \bibinfo {author}
  {\bibfnamefont {D.}~\bibnamefont {Pekker}}, \bibinfo {author} {\bibfnamefont
  {G.}~\bibnamefont {Refael}}, \bibinfo {author} {\bibfnamefont {J.~I.}\
  \bibnamefont {Cirac}}, \bibinfo {author} {\bibfnamefont {E.}~\bibnamefont
  {Demler}}, \bibinfo {author} {\bibfnamefont {M.~D.}\ \bibnamefont {Lukin}},\
  and\ \bibinfo {author} {\bibfnamefont {P.}~\bibnamefont {Zoller}},\
  }\bibfield  {title} {\bibinfo {title} {Majorana {{Fermions}} in
  {{Equilibrium}} and in {{Driven Cold-Atom Quantum Wires}}},\ }\href
  {https://doi.org/10.1103/PhysRevLett.106.220402} {\bibfield  {journal}
  {\bibinfo  {journal} {Physical Review Letters}\ }\textbf {\bibinfo {volume}
  {106}},\ \bibinfo {pages} {220402} (\bibinfo {year} {2011})}\BibitemShut
  {NoStop}%
\bibitem [{\citenamefont {Wang}\ \emph {et~al.}(2013)\citenamefont {Wang},
  \citenamefont {Steinberg}, \citenamefont {{Jarillo-Herrero}},\ and\
  \citenamefont {Gedik}}]{wangObservationFloquetBlochStates2013}%
  \BibitemOpen
  \bibfield  {author} {\bibinfo {author} {\bibfnamefont {Y.~H.}\ \bibnamefont
  {Wang}}, \bibinfo {author} {\bibfnamefont {H.}~\bibnamefont {Steinberg}},
  \bibinfo {author} {\bibfnamefont {P.}~\bibnamefont {{Jarillo-Herrero}}},\
  and\ \bibinfo {author} {\bibfnamefont {N.}~\bibnamefont {Gedik}},\ }\bibfield
   {title} {\bibinfo {title} {Observation of {{Floquet-Bloch States}} on the
  {{Surface}} of a {{Topological Insulator}}},\ }\href
  {https://doi.org/10.1126/science.1239834} {\bibfield  {journal} {\bibinfo
  {journal} {Science}\ }\textbf {\bibinfo {volume} {342}},\ \bibinfo {pages}
  {453} (\bibinfo {year} {2013})}\BibitemShut {NoStop}%
\bibitem [{\citenamefont {H{\"u}bener}\ \emph {et~al.}(2017)\citenamefont
  {H{\"u}bener}, \citenamefont {Sentef}, \citenamefont {De~Giovannini},
  \citenamefont {Kemper},\ and\ \citenamefont
  {Rubio}}]{hubenerCreatingStableFloquet2017}%
  \BibitemOpen
  \bibfield  {author} {\bibinfo {author} {\bibfnamefont {H.}~\bibnamefont
  {H{\"u}bener}}, \bibinfo {author} {\bibfnamefont {M.~A.}\ \bibnamefont
  {Sentef}}, \bibinfo {author} {\bibfnamefont {U.}~\bibnamefont
  {De~Giovannini}}, \bibinfo {author} {\bibfnamefont {A.~F.}\ \bibnamefont
  {Kemper}},\ and\ \bibinfo {author} {\bibfnamefont {A.}~\bibnamefont
  {Rubio}},\ }\bibfield  {title} {\bibinfo {title} {Creating stable
  {{Floquet}}--{{Weyl}} semimetals by laser-driving of {{3D Dirac}}
  materials},\ }\href {https://doi.org/10.1038/ncomms13940} {\bibfield
  {journal} {\bibinfo  {journal} {Nature Communications}\ }\textbf {\bibinfo
  {volume} {8}},\ \bibinfo {pages} {13940} (\bibinfo {year}
  {2017})}\BibitemShut {NoStop}%
\bibitem [{\citenamefont {Ito}\ \emph {et~al.}(2023)\citenamefont {Ito},
  \citenamefont {Sch{\"u}ler}, \citenamefont {Meierhofer}, \citenamefont
  {Schlauderer}, \citenamefont {Freudenstein}, \citenamefont {Reimann},
  \citenamefont {Afanasiev}, \citenamefont {Kokh}, \citenamefont
  {Tereshchenko}, \citenamefont {G{\"u}dde}, \citenamefont {Sentef},
  \citenamefont {H{\"o}fer},\ and\ \citenamefont
  {Huber}}]{itoBuildupDephasingFloquet2023}%
  \BibitemOpen
  \bibfield  {author} {\bibinfo {author} {\bibfnamefont {S.}~\bibnamefont
  {Ito}}, \bibinfo {author} {\bibfnamefont {M.}~\bibnamefont {Sch{\"u}ler}},
  \bibinfo {author} {\bibfnamefont {M.}~\bibnamefont {Meierhofer}}, \bibinfo
  {author} {\bibfnamefont {S.}~\bibnamefont {Schlauderer}}, \bibinfo {author}
  {\bibfnamefont {J.}~\bibnamefont {Freudenstein}}, \bibinfo {author}
  {\bibfnamefont {J.}~\bibnamefont {Reimann}}, \bibinfo {author} {\bibfnamefont
  {D.}~\bibnamefont {Afanasiev}}, \bibinfo {author} {\bibfnamefont {K.~A.}\
  \bibnamefont {Kokh}}, \bibinfo {author} {\bibfnamefont {O.~E.}\ \bibnamefont
  {Tereshchenko}}, \bibinfo {author} {\bibfnamefont {J.}~\bibnamefont
  {G{\"u}dde}}, \bibinfo {author} {\bibfnamefont {M.~A.}\ \bibnamefont
  {Sentef}}, \bibinfo {author} {\bibfnamefont {U.}~\bibnamefont {H{\"o}fer}},\
  and\ \bibinfo {author} {\bibfnamefont {R.}~\bibnamefont {Huber}},\ }\bibfield
   {title} {\bibinfo {title} {Build-up and dephasing of {{Floquet}}--{{Bloch}}
  bands on subcycle timescales},\ }\href
  {https://doi.org/10.1038/s41586-023-05850-x} {\bibfield  {journal} {\bibinfo
  {journal} {Nature}\ }\textbf {\bibinfo {volume} {616}},\ \bibinfo {pages}
  {696} (\bibinfo {year} {2023})}\BibitemShut {NoStop}%
\bibitem [{\citenamefont {De~La~Torre}\ \emph {et~al.}(2021)\citenamefont
  {De~La~Torre}, \citenamefont {Kennes}, \citenamefont {Claassen},
  \citenamefont {Gerber}, \citenamefont {McIver},\ and\ \citenamefont
  {Sentef}}]{delatorreColloquiumNonthermalPathways2021}%
  \BibitemOpen
  \bibfield  {author} {\bibinfo {author} {\bibfnamefont {A.}~\bibnamefont
  {De~La~Torre}}, \bibinfo {author} {\bibfnamefont {D.~M.}\ \bibnamefont
  {Kennes}}, \bibinfo {author} {\bibfnamefont {M.}~\bibnamefont {Claassen}},
  \bibinfo {author} {\bibfnamefont {S.}~\bibnamefont {Gerber}}, \bibinfo
  {author} {\bibfnamefont {J.~W.}\ \bibnamefont {McIver}},\ and\ \bibinfo
  {author} {\bibfnamefont {M.~A.}\ \bibnamefont {Sentef}},\ }\bibfield  {title}
  {\bibinfo {title} {{\emph{Colloquium:}} {{Nonthermal}} pathways to ultrafast
  control in quantum materials},\ }\href
  {https://doi.org/10.1103/RevModPhys.93.041002} {\bibfield  {journal}
  {\bibinfo  {journal} {Reviews of Modern Physics}\ }\textbf {\bibinfo {volume}
  {93}},\ \bibinfo {pages} {041002} (\bibinfo {year} {2021})}\BibitemShut
  {NoStop}%
\bibitem [{\citenamefont {Beaurepaire}\ \emph {et~al.}(1996)\citenamefont
  {Beaurepaire}, \citenamefont {Merle}, \citenamefont {Daunois},\ and\
  \citenamefont {Bigot}}]{beaurepaireUltrafastSpinDynamics1996}%
  \BibitemOpen
  \bibfield  {author} {\bibinfo {author} {\bibfnamefont {E.}~\bibnamefont
  {Beaurepaire}}, \bibinfo {author} {\bibfnamefont {J.-C.}\ \bibnamefont
  {Merle}}, \bibinfo {author} {\bibfnamefont {A.}~\bibnamefont {Daunois}},\
  and\ \bibinfo {author} {\bibfnamefont {J.-Y.}\ \bibnamefont {Bigot}},\
  }\bibfield  {title} {\bibinfo {title} {Ultrafast {{Spin Dynamics}} in
  {{Ferromagnetic Nickel}}},\ }\href
  {https://doi.org/10.1103/PhysRevLett.76.4250} {\bibfield  {journal} {\bibinfo
   {journal} {Physical Review Letters}\ }\textbf {\bibinfo {volume} {76}},\
  \bibinfo {pages} {4250} (\bibinfo {year} {1996})}\BibitemShut {NoStop}%
\bibitem [{\citenamefont {Koopmans}\ \emph {et~al.}(2000)\citenamefont
  {Koopmans}, \citenamefont {Van~Kampen}, \citenamefont {Kohlhepp},\ and\
  \citenamefont {De~Jonge}}]{koopmansUltrafastMagnetoOpticsNickel2000}%
  \BibitemOpen
  \bibfield  {author} {\bibinfo {author} {\bibfnamefont {B.}~\bibnamefont
  {Koopmans}}, \bibinfo {author} {\bibfnamefont {M.}~\bibnamefont
  {Van~Kampen}}, \bibinfo {author} {\bibfnamefont {J.~T.}\ \bibnamefont
  {Kohlhepp}},\ and\ \bibinfo {author} {\bibfnamefont {W.~J.~M.}\ \bibnamefont
  {De~Jonge}},\ }\bibfield  {title} {\bibinfo {title} {Ultrafast
  {{Magneto-Optics}} in {{Nickel}}: {{Magnetism}} or {{Optics}}?},\ }\href
  {https://doi.org/10.1103/PhysRevLett.85.844} {\bibfield  {journal} {\bibinfo
  {journal} {Physical Review Letters}\ }\textbf {\bibinfo {volume} {85}},\
  \bibinfo {pages} {844} (\bibinfo {year} {2000})}\BibitemShut {NoStop}%
\bibitem [{\citenamefont {Eschenlohr}\ \emph {et~al.}(2013)\citenamefont
  {Eschenlohr}, \citenamefont {Battiato}, \citenamefont {Maldonado},
  \citenamefont {Pontius}, \citenamefont {Kachel}, \citenamefont {Holldack},
  \citenamefont {Mitzner}, \citenamefont {F{\"o}hlisch}, \citenamefont
  {Oppeneer},\ and\ \citenamefont
  {Stamm}}]{eschenlohrUltrafastSpinTransport2013}%
  \BibitemOpen
  \bibfield  {author} {\bibinfo {author} {\bibfnamefont {A.}~\bibnamefont
  {Eschenlohr}}, \bibinfo {author} {\bibfnamefont {M.}~\bibnamefont
  {Battiato}}, \bibinfo {author} {\bibfnamefont {P.}~\bibnamefont {Maldonado}},
  \bibinfo {author} {\bibfnamefont {N.}~\bibnamefont {Pontius}}, \bibinfo
  {author} {\bibfnamefont {T.}~\bibnamefont {Kachel}}, \bibinfo {author}
  {\bibfnamefont {K.}~\bibnamefont {Holldack}}, \bibinfo {author}
  {\bibfnamefont {R.}~\bibnamefont {Mitzner}}, \bibinfo {author} {\bibfnamefont
  {A.}~\bibnamefont {F{\"o}hlisch}}, \bibinfo {author} {\bibfnamefont {P.~M.}\
  \bibnamefont {Oppeneer}},\ and\ \bibinfo {author} {\bibfnamefont
  {C.}~\bibnamefont {Stamm}},\ }\bibfield  {title} {\bibinfo {title} {Ultrafast
  spin transport as key to femtosecond demagnetization},\ }\href
  {https://doi.org/10.1038/nmat3546} {\bibfield  {journal} {\bibinfo  {journal}
  {Nature Materials}\ }\textbf {\bibinfo {volume} {12}},\ \bibinfo {pages}
  {332} (\bibinfo {year} {2013})}\BibitemShut {NoStop}%
\bibitem [{\citenamefont {Waldecker}\ \emph {et~al.}(2017)\citenamefont
  {Waldecker}, \citenamefont {Bertoni}, \citenamefont {H{\"u}bener},
  \citenamefont {Brumme}, \citenamefont {Vasileiadis}, \citenamefont {Zahn},
  \citenamefont {Rubio},\ and\ \citenamefont
  {Ernstorfer}}]{waldeckerMomentumResolvedViewElectronPhonon2017}%
  \BibitemOpen
  \bibfield  {author} {\bibinfo {author} {\bibfnamefont {L.}~\bibnamefont
  {Waldecker}}, \bibinfo {author} {\bibfnamefont {R.}~\bibnamefont {Bertoni}},
  \bibinfo {author} {\bibfnamefont {H.}~\bibnamefont {H{\"u}bener}}, \bibinfo
  {author} {\bibfnamefont {T.}~\bibnamefont {Brumme}}, \bibinfo {author}
  {\bibfnamefont {T.}~\bibnamefont {Vasileiadis}}, \bibinfo {author}
  {\bibfnamefont {D.}~\bibnamefont {Zahn}}, \bibinfo {author} {\bibfnamefont
  {A.}~\bibnamefont {Rubio}},\ and\ \bibinfo {author} {\bibfnamefont
  {R.}~\bibnamefont {Ernstorfer}},\ }\bibfield  {title} {\bibinfo {title}
  {Momentum-{{Resolved View}} of {{Electron-Phonon Coupling}} in {{Multilayer
  WSe}} 2},\ }\href {https://doi.org/10.1103/PhysRevLett.119.036803} {\bibfield
   {journal} {\bibinfo  {journal} {Physical Review Letters}\ }\textbf {\bibinfo
  {volume} {119}},\ \bibinfo {pages} {036803} (\bibinfo {year}
  {2017})}\BibitemShut {NoStop}%
\bibitem [{\citenamefont {Shallcross}\ \emph {et~al.}(2022)\citenamefont
  {Shallcross}, \citenamefont {Li}, \citenamefont {Dewhurst}, \citenamefont
  {Sharma},\ and\ \citenamefont
  {Elliott}}]{shallcrossUltrafastOpticalControl2022}%
  \BibitemOpen
  \bibfield  {author} {\bibinfo {author} {\bibfnamefont {S.}~\bibnamefont
  {Shallcross}}, \bibinfo {author} {\bibfnamefont {Q.~Z.}\ \bibnamefont {Li}},
  \bibinfo {author} {\bibfnamefont {J.~K.}\ \bibnamefont {Dewhurst}}, \bibinfo
  {author} {\bibfnamefont {S.}~\bibnamefont {Sharma}},\ and\ \bibinfo {author}
  {\bibfnamefont {P.}~\bibnamefont {Elliott}},\ }\bibfield  {title} {\bibinfo
  {title} {Ultrafast optical control over spin and momentum in solids},\ }\href
  {https://doi.org/10.1063/5.0076198} {\bibfield  {journal} {\bibinfo
  {journal} {Applied Physics Letters}\ }\textbf {\bibinfo {volume} {120}},\
  \bibinfo {pages} {032403} (\bibinfo {year} {2022})}\BibitemShut {NoStop}%
\bibitem [{\citenamefont {Sharma}\ \emph {et~al.}(2022)\citenamefont {Sharma},
  \citenamefont {Elliott},\ and\ \citenamefont
  {Shallcross}}]{sharmaValleyControlLinearly2022}%
  \BibitemOpen
  \bibfield  {author} {\bibinfo {author} {\bibfnamefont {S.}~\bibnamefont
  {Sharma}}, \bibinfo {author} {\bibfnamefont {P.}~\bibnamefont {Elliott}},\
  and\ \bibinfo {author} {\bibfnamefont {S.}~\bibnamefont {Shallcross}},\
  }\bibfield  {title} {\bibinfo {title} {Valley control by linearly polarized
  laser pulses: Example of {{WSe}} {\textsubscript{2}}},\ }\href
  {https://doi.org/10.1364/OPTICA.458991} {\bibfield  {journal} {\bibinfo
  {journal} {Optica}\ }\textbf {\bibinfo {volume} {9}},\ \bibinfo {pages} {947}
  (\bibinfo {year} {2022})}\BibitemShut {NoStop}%
\bibitem [{\citenamefont {Sch{\"u}ler}\ \emph {et~al.}(2020)\citenamefont
  {Sch{\"u}ler}, \citenamefont {De~Giovannini}, \citenamefont {H{\"u}bener},
  \citenamefont {Rubio}, \citenamefont {Sentef},\ and\ \citenamefont
  {Werner}}]{schulerLocalBerryCurvature2020}%
  \BibitemOpen
  \bibfield  {author} {\bibinfo {author} {\bibfnamefont {M.}~\bibnamefont
  {Sch{\"u}ler}}, \bibinfo {author} {\bibfnamefont {U.}~\bibnamefont
  {De~Giovannini}}, \bibinfo {author} {\bibfnamefont {H.}~\bibnamefont
  {H{\"u}bener}}, \bibinfo {author} {\bibfnamefont {A.}~\bibnamefont {Rubio}},
  \bibinfo {author} {\bibfnamefont {M.~A.}\ \bibnamefont {Sentef}},\ and\
  \bibinfo {author} {\bibfnamefont {P.}~\bibnamefont {Werner}},\ }\bibfield
  {title} {\bibinfo {title} {Local {{Berry}} curvature signatures in dichroic
  angle-resolved photoelectron spectroscopy from two-dimensional materials},\
  }\href {https://doi.org/10.1126/sciadv.aay2730} {\bibfield  {journal}
  {\bibinfo  {journal} {Science Advances}\ }\textbf {\bibinfo {volume} {6}},\
  \bibinfo {pages} {eaay2730} (\bibinfo {year} {2020})}\BibitemShut {NoStop}%
\bibitem [{\citenamefont {Alliati}\ and\ \citenamefont
  {Gr{\"u}ning}(2023)}]{alliatiFloquetFormulationDynamical2023}%
  \BibitemOpen
  \bibfield  {author} {\bibinfo {author} {\bibfnamefont {I.~M.}\ \bibnamefont
  {Alliati}}\ and\ \bibinfo {author} {\bibfnamefont {M.}~\bibnamefont
  {Gr{\"u}ning}},\ }\bibfield  {title} {\bibinfo {title} {Floquet formulation
  of the dynamical {{Berry-phase}} approach to nonlinear optics in extended
  systems},\ }\href {https://doi.org/10.1088/2516-1075/acbc5e} {\bibfield
  {journal} {\bibinfo  {journal} {Electronic Structure}\ }\textbf {\bibinfo
  {volume} {5}},\ \bibinfo {pages} {017001} (\bibinfo {year}
  {2023})}\BibitemShut {NoStop}%
\bibitem [{\citenamefont
  {Holthaus}(2016)}]{holthausFloquetEngineeringQuasienergy2016}%
  \BibitemOpen
  \bibfield  {author} {\bibinfo {author} {\bibfnamefont {M.}~\bibnamefont
  {Holthaus}},\ }\bibfield  {title} {\bibinfo {title} {Floquet engineering with
  quasienergy bands of periodically driven optical lattices},\ }\href
  {https://doi.org/10.1088/0953-4075/49/1/013001} {\bibfield  {journal}
  {\bibinfo  {journal} {Journal of Physics B: Atomic, Molecular and Optical
  Physics}\ }\textbf {\bibinfo {volume} {49}},\ \bibinfo {pages} {013001}
  (\bibinfo {year} {2016})}\BibitemShut {NoStop}%
\bibitem [{\citenamefont {Blount}(1962)}]{blountFormalismsBandTheory1962}%
  \BibitemOpen
  \bibfield  {author} {\bibinfo {author} {\bibfnamefont {E.}~\bibnamefont
  {Blount}},\ }\bibfield  {title} {\bibinfo {title} {Formalisms of {{Band
  Theory}}},\ }in\ \href {https://doi.org/10.1016/S0081-1947(08)60459-2} {\emph
  {\bibinfo {booktitle} {Solid {{State Physics}}}}},\ Vol.~\bibinfo {volume}
  {13}\ (\bibinfo  {publisher} {Elsevier},\ \bibinfo {address} {New York \&
  London},\ \bibinfo {year} {1962})\ pp.\ \bibinfo {pages}
  {305--373}\BibitemShut {NoStop}%
\bibitem [{\citenamefont {{Cohen-Tannoudji}}\ \emph {et~al.}(1977)\citenamefont
  {{Cohen-Tannoudji}}, \citenamefont {Diu},\ and\ \citenamefont
  {Lalo{\"e}}}]{cohen-tannoudjiQuantumMechanics1977}%
  \BibitemOpen
  \bibfield  {author} {\bibinfo {author} {\bibfnamefont {C.}~\bibnamefont
  {{Cohen-Tannoudji}}}, \bibinfo {author} {\bibfnamefont {B.}~\bibnamefont
  {Diu}},\ and\ \bibinfo {author} {\bibfnamefont {F.}~\bibnamefont
  {Lalo{\"e}}},\ }\href@noop {} {\emph {\bibinfo {title} {{Quantum
  mechanics}}}}\ (\bibinfo  {publisher} {Wiley},\ \bibinfo {address} {New
  York},\ \bibinfo {year} {1977})\BibitemShut {NoStop}%
\bibitem [{\citenamefont {Sakurai}\ and\ \citenamefont
  {Napolitano}(2017)}]{sakuraiModernQuantumMechanics2017}%
  \BibitemOpen
  \bibfield  {author} {\bibinfo {author} {\bibfnamefont {J.~J.}\ \bibnamefont
  {Sakurai}}\ and\ \bibinfo {author} {\bibfnamefont {J.}~\bibnamefont
  {Napolitano}},\ }\href {https://doi.org/10.1017/9781108499996} {\emph
  {\bibinfo {title} {Modern {{Quantum Mechanics}}}}},\ \bibinfo {edition}
  {2nd}\ ed.\ (\bibinfo  {publisher} {Cambridge University Press},\ \bibinfo
  {address} {Cambridge},\ \bibinfo {year} {2017})\BibitemShut {NoStop}%
\bibitem [{\citenamefont {Kobe}\ and\ \citenamefont
  {Yang}(1985)}]{kobeGaugeTransformationTimeevolution1985}%
  \BibitemOpen
  \bibfield  {author} {\bibinfo {author} {\bibfnamefont {D.~H.}\ \bibnamefont
  {Kobe}}\ and\ \bibinfo {author} {\bibfnamefont {K.-H.}\ \bibnamefont
  {Yang}},\ }\bibfield  {title} {\bibinfo {title} {Gauge transformation of the
  time-evolution operator},\ }\href {https://doi.org/10.1103/PhysRevA.32.952}
  {\bibfield  {journal} {\bibinfo  {journal} {Physical Review A}\ }\textbf
  {\bibinfo {volume} {32}},\ \bibinfo {pages} {952} (\bibinfo {year}
  {1985})}\BibitemShut {NoStop}%
\bibitem [{Note1()}]{Note1}%
  \BibitemOpen
  \bibinfo {note} {We first remark that all unitary operators are normal, so
  the spectral theorem applies. This means that $\protect \hat {Q}$ can be
  decomposed as $\protect \hat {U}\protect \hat {D}\protect \hat {U}^{\dagger
  }$, where $\protect \hat {U}$ is unitary, and $\protect \hat {D}$ diagonal.
  Since $\protect \hat {Q}$ is unitary, it is trivial to show that its
  eigenvalues, the entries of $\protect \hat {D}$, are complex numbers in the
  complex unit circle. With this remark in place, we argue that $\protect \hat
  {W}\protect \hat {Q}\protect \hat {W}^{\dagger }$ is again unitary with
  eigenvalues in the complex unit circle. This guarantees that the Taylor
  expansion for the complex logarithm is convergent. Lastly, some algebra shows
  that \begin {equation} \begin {split} &\protect \text {log}(\protect \hat
  {W}\protect \hat {Q}\protect \hat {W}^{\dagger }) = \DOTSB \sum@ \slimits@
  _{k=1}^{\infty }\protect \frac {(-1)^{k+1}}{k}\left (\protect \hat
  {W}\protect \hat {Q}\protect \hat {W}^{\dagger } - \protect \hat {\protect
  \mathds {1}}\right )^k = \\ & \DOTSB \sum@ \slimits@ _{k=1}^{\infty }\protect
  \frac {(-1)^{k+1}}{k}\left (\protect \hat {W}\protect \hat {Q}\protect \hat
  {W}^{\dagger } - \protect \hat {W}\protect \hat {W}^{\dagger }\right )^k = \\
  &\protect \hat {W}\left [\DOTSB \sum@ \slimits@ _{k=1}^{\infty }\protect
  \frac {(-1)^{k+1}}{k}\left (\protect \hat {Q} - \protect \hat {\protect
  \mathds {1}}\right )^k\right ]\protect \hat {W}^{\dagger } = \protect \hat
  {W}\protect \text {log}(\protect \hat {Q})\protect \hat {W}^{\dagger } \end
  {split} \end {equation} where we have made extensive use of the unitarity
  property of $\protect \hat {W}$.}\BibitemShut {Stop}%
\bibitem [{\citenamefont {Passos}\ \emph {et~al.}(2018)\citenamefont {Passos},
  \citenamefont {Ventura}, \citenamefont {Lopes}, \citenamefont {Santos},\ and\
  \citenamefont {Peres}}]{passosNonlinearOpticalResponses2018}%
  \BibitemOpen
  \bibfield  {author} {\bibinfo {author} {\bibfnamefont {D.~J.}\ \bibnamefont
  {Passos}}, \bibinfo {author} {\bibfnamefont {G.~B.}\ \bibnamefont {Ventura}},
  \bibinfo {author} {\bibfnamefont {J.~M. V.~P.}\ \bibnamefont {Lopes}},
  \bibinfo {author} {\bibfnamefont {J.~M. B. L.~D.}\ \bibnamefont {Santos}},\
  and\ \bibinfo {author} {\bibfnamefont {N.~M.~R.}\ \bibnamefont {Peres}},\
  }\bibfield  {title} {\bibinfo {title} {Nonlinear optical responses of
  crystalline systems: {{Results}} from a velocity gauge analysis},\ }\href
  {https://doi.org/10.1103/PhysRevB.97.235446} {\bibfield  {journal} {\bibinfo
  {journal} {Physical Review B}\ }\textbf {\bibinfo {volume} {97}},\ \bibinfo
  {pages} {235446} (\bibinfo {year} {2018})}\BibitemShut {NoStop}%
\bibitem [{Note2()}]{Note2}%
  \BibitemOpen
  \bibinfo {note} {In the context of atomic physics, this gauge is more
  commonly referred as the radiation gauge. The Hamiltonian in the velocity
  gauge is given by $\protect \hat {H}_0 - \protect \frac {q}{m}\protect \bm
  {A}(t)\cdot \protect \hat {\protect \bm {p}}$. Both gauges differ only by a
  time dependent operator multiple of the identity, which is irrelevant to the
  dynamics.}\BibitemShut {Stop}%
\bibitem [{\citenamefont {Scully}\ and\ \citenamefont
  {Zubairy}(1997)}]{scullyQuantumOptics1997}%
  \BibitemOpen
  \bibfield  {author} {\bibinfo {author} {\bibfnamefont {M.~O.}\ \bibnamefont
  {Scully}}\ and\ \bibinfo {author} {\bibfnamefont {M.~S.}\ \bibnamefont
  {Zubairy}},\ }\href {https://doi.org/10.1017/CBO9780511813993} {\emph
  {\bibinfo {title} {Quantum {{Optics}}}}},\ \bibinfo {edition} {1st}\ ed.\
  (\bibinfo  {publisher} {Cambridge University Press},\ \bibinfo {address}
  {Cambridge},\ \bibinfo {year} {1997})\BibitemShut {NoStop}%
\bibitem [{\citenamefont {Ventura}\ \emph {et~al.}(2017)\citenamefont
  {Ventura}, \citenamefont {Passos}, \citenamefont {{Lopes dos Santos}},
  \citenamefont {Viana Parente~Lopes},\ and\ \citenamefont
  {Peres}}]{venturaGaugeCovariancesNonlinear2017}%
  \BibitemOpen
  \bibfield  {author} {\bibinfo {author} {\bibfnamefont {G.~B.}\ \bibnamefont
  {Ventura}}, \bibinfo {author} {\bibfnamefont {D.~J.}\ \bibnamefont {Passos}},
  \bibinfo {author} {\bibfnamefont {J.~M.~B.}\ \bibnamefont {{Lopes dos
  Santos}}}, \bibinfo {author} {\bibfnamefont {J.~M.}\ \bibnamefont {Viana
  Parente~Lopes}},\ and\ \bibinfo {author} {\bibfnamefont {N.~M.~R.}\
  \bibnamefont {Peres}},\ }\bibfield  {title} {\bibinfo {title} {Gauge
  covariances and nonlinear optical responses},\ }\href
  {https://doi.org/10.1103/PhysRevB.96.035431} {\bibfield  {journal} {\bibinfo
  {journal} {Physical Review B}\ }\textbf {\bibinfo {volume} {96}},\ \bibinfo
  {pages} {035431} (\bibinfo {year} {2017})}\BibitemShut {NoStop}%
\bibitem [{\citenamefont {Evarestov}\ and\ \citenamefont
  {Smirnov}(1997)}]{evarestovSiteSymmetryCrystals1997}%
  \BibitemOpen
  \bibfield  {author} {\bibinfo {author} {\bibfnamefont {R.~A.}\ \bibnamefont
  {Evarestov}}\ and\ \bibinfo {author} {\bibfnamefont {V.~P.}\ \bibnamefont
  {Smirnov}},\ }\href {https://doi.org/10.1007/978-3-642-60488-1} {\emph
  {\bibinfo {title} {Site {{Symmetry}} in {{Crystals}}}}},\ edited by\ \bibinfo
  {editor} {\bibfnamefont {M.}~\bibnamefont {Cardona}}, \bibinfo {editor}
  {\bibfnamefont {P.}~\bibnamefont {Fulde}}, \bibinfo {editor} {\bibfnamefont
  {K.}~\bibnamefont {{von Klitzing}}},\ and\ \bibinfo {editor} {\bibfnamefont
  {H.-J.}\ \bibnamefont {Queisser}},\ \bibinfo {series} {Springer {{Series}} in
  {{Solid-State Sciences}}}, Vol.\ \bibinfo {volume} {108}\ (\bibinfo
  {publisher} {Springer Berlin Heidelberg},\ \bibinfo {address} {Berlin,
  Heidelberg},\ \bibinfo {year} {1997})\BibitemShut {NoStop}%
\bibitem [{\citenamefont {Bohm}\ \emph {et~al.}(2003)\citenamefont {Bohm},
  \citenamefont {Mostafazadeh}, \citenamefont {Koizumi}, \citenamefont {Niu},\
  and\ \citenamefont {Zwanziger}}]{bohmGeometricPhaseQuantum2003}%
  \BibitemOpen
  \bibfield  {author} {\bibinfo {author} {\bibfnamefont {A.}~\bibnamefont
  {Bohm}}, \bibinfo {author} {\bibfnamefont {A.}~\bibnamefont {Mostafazadeh}},
  \bibinfo {author} {\bibfnamefont {H.}~\bibnamefont {Koizumi}}, \bibinfo
  {author} {\bibfnamefont {Q.}~\bibnamefont {Niu}},\ and\ \bibinfo {author}
  {\bibfnamefont {J.}~\bibnamefont {Zwanziger}},\ }\href
  {https://doi.org/10.1007/978-3-662-10333-3} {\emph {\bibinfo {title} {The
  {{Geometric Phase}} in {{Quantum Systems}}}}}\ (\bibinfo  {publisher}
  {Springer Berlin Heidelberg},\ \bibinfo {address} {Berlin, Heidelberg},\
  \bibinfo {year} {2003})\BibitemShut {NoStop}%
\bibitem [{\citenamefont {Wang}\ \emph {et~al.}(2006)\citenamefont {Wang},
  \citenamefont {Yates}, \citenamefont {Souza},\ and\ \citenamefont
  {Vanderbilt}}]{wangInitioCalculationAnomalous2006}%
  \BibitemOpen
  \bibfield  {author} {\bibinfo {author} {\bibfnamefont {X.}~\bibnamefont
  {Wang}}, \bibinfo {author} {\bibfnamefont {J.~R.}\ \bibnamefont {Yates}},
  \bibinfo {author} {\bibfnamefont {I.}~\bibnamefont {Souza}},\ and\ \bibinfo
  {author} {\bibfnamefont {D.}~\bibnamefont {Vanderbilt}},\ }\bibfield  {title}
  {\bibinfo {title} {{\emph{Ab Initio}} calculation of the anomalous {{Hall}}
  conductivity by {{Wannier}} interpolation},\ }\href
  {https://doi.org/10.1103/PhysRevB.74.195118} {\bibfield  {journal} {\bibinfo
  {journal} {Physical Review B}\ }\textbf {\bibinfo {volume} {74}},\ \bibinfo
  {pages} {195118} (\bibinfo {year} {2006})}\BibitemShut {NoStop}%
\bibitem [{\citenamefont
  {Vanderbilt}(2018)}]{vanderbiltBerryPhasesElectronic2018}%
  \BibitemOpen
  \bibfield  {author} {\bibinfo {author} {\bibfnamefont {D.}~\bibnamefont
  {Vanderbilt}},\ }\href {https://doi.org/10.1017/9781316662205} {\emph
  {\bibinfo {title} {Berry {{Phases}} in {{Electronic Structure Theory}}:
  {{Electric Polarization}}, {{Orbital Magnetization}} and {{Topological
  Insulators}}}}},\ \bibinfo {edition} {1st}\ ed.\ (\bibinfo  {publisher}
  {Cambridge University Press},\ \bibinfo {address} {Cambridge},\ \bibinfo
  {year} {2018})\BibitemShut {NoStop}%
\bibitem [{\citenamefont {Parker}\ \emph {et~al.}(2019)\citenamefont {Parker},
  \citenamefont {Morimoto}, \citenamefont {Orenstein},\ and\ \citenamefont
  {Moore}}]{parkerDiagrammaticApproachNonlinear2019}%
  \BibitemOpen
  \bibfield  {author} {\bibinfo {author} {\bibfnamefont {D.~E.}\ \bibnamefont
  {Parker}}, \bibinfo {author} {\bibfnamefont {T.}~\bibnamefont {Morimoto}},
  \bibinfo {author} {\bibfnamefont {J.}~\bibnamefont {Orenstein}},\ and\
  \bibinfo {author} {\bibfnamefont {J.~E.}\ \bibnamefont {Moore}},\ }\bibfield
  {title} {\bibinfo {title} {Diagrammatic approach to nonlinear optical
  response with application to {{Weyl}} semimetals},\ }\href
  {https://doi.org/10.1103/PhysRevB.99.045121} {\bibfield  {journal} {\bibinfo
  {journal} {Physical Review B}\ }\textbf {\bibinfo {volume} {99}},\ \bibinfo
  {pages} {045121} (\bibinfo {year} {2019})}\BibitemShut {NoStop}%
\bibitem [{\citenamefont {Aversa}\ and\ \citenamefont
  {Sipe}(1995)}]{aversaNonlinearOpticalSusceptibilities1995}%
  \BibitemOpen
  \bibfield  {author} {\bibinfo {author} {\bibfnamefont {C.}~\bibnamefont
  {Aversa}}\ and\ \bibinfo {author} {\bibfnamefont {J.~E.}\ \bibnamefont
  {Sipe}},\ }\bibfield  {title} {\bibinfo {title} {Nonlinear optical
  susceptibilities of semiconductors: {{Results}} with a length-gauge
  analysis},\ }\href {https://doi.org/10.1103/PhysRevB.52.14636} {\bibfield
  {journal} {\bibinfo  {journal} {Physical Review B}\ }\textbf {\bibinfo
  {volume} {52}},\ \bibinfo {pages} {14636} (\bibinfo {year}
  {1995})}\BibitemShut {NoStop}%
\bibitem [{\citenamefont
  {Holthaus}(1992{\natexlab{a}})}]{holthausQuantumTheoryIdeal1992}%
  \BibitemOpen
  \bibfield  {author} {\bibinfo {author} {\bibfnamefont {M.}~\bibnamefont
  {Holthaus}},\ }\bibfield  {title} {\bibinfo {title} {The quantum theory of an
  ideal superlattice responding to far-infrared laser radiation},\ }\href
  {https://doi.org/10.1007/BF01320944} {\bibfield  {journal} {\bibinfo
  {journal} {Z. Physik B - Condensed Matter}\ }\textbf {\bibinfo {volume}
  {89}},\ \bibinfo {pages} {251} (\bibinfo {year}
  {1992}{\natexlab{a}})}\BibitemShut {NoStop}%
\bibitem [{\citenamefont
  {Holthaus}(1992{\natexlab{b}})}]{holthausCollapseMinibandsFarinfrared1992}%
  \BibitemOpen
  \bibfield  {author} {\bibinfo {author} {\bibfnamefont {M.}~\bibnamefont
  {Holthaus}},\ }\bibfield  {title} {\bibinfo {title} {Collapse of minibands in
  far-infrared irradiated superlattices},\ }\href
  {https://doi.org/10.1103/PhysRevLett.69.351} {\bibfield  {journal} {\bibinfo
  {journal} {Physical Review Letters}\ }\textbf {\bibinfo {volume} {69}},\
  \bibinfo {pages} {351} (\bibinfo {year} {1992}{\natexlab{b}})}\BibitemShut
  {NoStop}%
\bibitem [{\citenamefont {Hone}\ and\ \citenamefont
  {Holthaus}(1993)}]{honeLocallyDisorderedLattices1993}%
  \BibitemOpen
  \bibfield  {author} {\bibinfo {author} {\bibfnamefont {D.~W.}\ \bibnamefont
  {Hone}}\ and\ \bibinfo {author} {\bibfnamefont {M.}~\bibnamefont
  {Holthaus}},\ }\bibfield  {title} {\bibinfo {title} {Locally disordered
  lattices in strong ac electric fields},\ }\href
  {https://doi.org/10.1103/PhysRevB.48.15123} {\bibfield  {journal} {\bibinfo
  {journal} {Physical Review B}\ }\textbf {\bibinfo {volume} {48}},\ \bibinfo
  {pages} {15123} (\bibinfo {year} {1993})}\BibitemShut {NoStop}%
\bibitem [{\citenamefont {Pan}\ \emph {et~al.}(2006)\citenamefont {Pan},
  \citenamefont {Sun},\ and\ \citenamefont
  {Chen}}]{panInterlayerStackingNature2006}%
  \BibitemOpen
  \bibfield  {author} {\bibinfo {author} {\bibfnamefont {Z.}~\bibnamefont
  {Pan}}, \bibinfo {author} {\bibfnamefont {H.}~\bibnamefont {Sun}},\ and\
  \bibinfo {author} {\bibfnamefont {C.}~\bibnamefont {Chen}},\ }\bibfield
  {title} {\bibinfo {title} {Interlayer stacking and nature of the electronic
  band gap in graphitic {{B C}} 2 {{N}} : {{First-principles}} pseudopotential
  calculations},\ }\href {https://doi.org/10.1103/PhysRevB.73.193304}
  {\bibfield  {journal} {\bibinfo  {journal} {Physical Review B}\ }\textbf
  {\bibinfo {volume} {73}},\ \bibinfo {pages} {193304} (\bibinfo {year}
  {2006})}\BibitemShut {NoStop}%
\bibitem [{\citenamefont {{Iba{\~n}ez-Azpiroz}}\ \emph
  {et~al.}(2020)\citenamefont {{Iba{\~n}ez-Azpiroz}}, \citenamefont {Souza},\
  and\ \citenamefont {{de Juan}}}]{ibanez-azpirozDirectionalShiftCurrent2020}%
  \BibitemOpen
  \bibfield  {author} {\bibinfo {author} {\bibfnamefont {J.}~\bibnamefont
  {{Iba{\~n}ez-Azpiroz}}}, \bibinfo {author} {\bibfnamefont {I.}~\bibnamefont
  {Souza}},\ and\ \bibinfo {author} {\bibfnamefont {F.}~\bibnamefont {{de
  Juan}}},\ }\bibfield  {title} {\bibinfo {title} {Directional shift current in
  mirror-symmetric {{BC}} 2 {{N}}},\ }\href
  {https://doi.org/10.1103/PhysRevResearch.2.013263} {\bibfield  {journal}
  {\bibinfo  {journal} {Physical Review Research}\ }\textbf {\bibinfo {volume}
  {2}},\ \bibinfo {pages} {013263} (\bibinfo {year} {2020})}\BibitemShut
  {NoStop}%
\bibitem [{\citenamefont {Giannozzi}\ \emph {et~al.}(2020)\citenamefont
  {Giannozzi}, \citenamefont {Baseggio}, \citenamefont {Bonf{\`a}},
  \citenamefont {Brunato}, \citenamefont {Car}, \citenamefont {Carnimeo},
  \citenamefont {Cavazzoni}, \citenamefont {De~Gironcoli}, \citenamefont
  {Delugas}, \citenamefont {Ferrari~Ruffino}, \citenamefont {Ferretti},
  \citenamefont {Marzari}, \citenamefont {Timrov}, \citenamefont {Urru},\ and\
  \citenamefont {Baroni}}]{giannozziUantumESPRESSOExascale2020}%
  \BibitemOpen
  \bibfield  {author} {\bibinfo {author} {\bibfnamefont {P.}~\bibnamefont
  {Giannozzi}}, \bibinfo {author} {\bibfnamefont {O.}~\bibnamefont {Baseggio}},
  \bibinfo {author} {\bibfnamefont {P.}~\bibnamefont {Bonf{\`a}}}, \bibinfo
  {author} {\bibfnamefont {D.}~\bibnamefont {Brunato}}, \bibinfo {author}
  {\bibfnamefont {R.}~\bibnamefont {Car}}, \bibinfo {author} {\bibfnamefont
  {I.}~\bibnamefont {Carnimeo}}, \bibinfo {author} {\bibfnamefont
  {C.}~\bibnamefont {Cavazzoni}}, \bibinfo {author} {\bibfnamefont
  {S.}~\bibnamefont {De~Gironcoli}}, \bibinfo {author} {\bibfnamefont
  {P.}~\bibnamefont {Delugas}}, \bibinfo {author} {\bibfnamefont
  {F.}~\bibnamefont {Ferrari~Ruffino}}, \bibinfo {author} {\bibfnamefont
  {A.}~\bibnamefont {Ferretti}}, \bibinfo {author} {\bibfnamefont
  {N.}~\bibnamefont {Marzari}}, \bibinfo {author} {\bibfnamefont
  {I.}~\bibnamefont {Timrov}}, \bibinfo {author} {\bibfnamefont
  {A.}~\bibnamefont {Urru}},\ and\ \bibinfo {author} {\bibfnamefont
  {S.}~\bibnamefont {Baroni}},\ }\bibfield  {title} {\bibinfo {title} {Q
  {\textsc{uantum}} {{ESPRESSO}} toward the exascale},\ }\href
  {https://doi.org/10.1063/5.0005082} {\bibfield  {journal} {\bibinfo
  {journal} {The Journal of Chemical Physics}\ }\textbf {\bibinfo {volume}
  {152}},\ \bibinfo {pages} {154105} (\bibinfo {year} {2020})}\BibitemShut
  {NoStop}%
\bibitem [{\citenamefont {Pizzi}\ \emph {et~al.}(2020)\citenamefont {Pizzi},
  \citenamefont {Vitale}, \citenamefont {Arita}, \citenamefont {Bl{\"u}gel},
  \citenamefont {Freimuth}, \citenamefont {G{\'e}ranton}, \citenamefont
  {Gibertini}, \citenamefont {Gresch}, \citenamefont {Johnson}, \citenamefont
  {Koretsune}, \citenamefont {{Iba{\~n}ez-Azpiroz}}, \citenamefont {Lee},
  \citenamefont {Lihm}, \citenamefont {Marchand}, \citenamefont {Marrazzo},
  \citenamefont {Mokrousov}, \citenamefont {Mustafa}, \citenamefont {Nohara},
  \citenamefont {Nomura}, \citenamefont {Paulatto}, \citenamefont {Ponc{\'e}},
  \citenamefont {Ponweiser}, \citenamefont {Qiao}, \citenamefont {Th{\"o}le},
  \citenamefont {Tsirkin}, \citenamefont {Wierzbowska}, \citenamefont
  {Marzari}, \citenamefont {Vanderbilt}, \citenamefont {Souza}, \citenamefont
  {Mostofi},\ and\ \citenamefont {Yates}}]{pizziWannier90CommunityCode2020}%
  \BibitemOpen
  \bibfield  {author} {\bibinfo {author} {\bibfnamefont {G.}~\bibnamefont
  {Pizzi}}, \bibinfo {author} {\bibfnamefont {V.}~\bibnamefont {Vitale}},
  \bibinfo {author} {\bibfnamefont {R.}~\bibnamefont {Arita}}, \bibinfo
  {author} {\bibfnamefont {S.}~\bibnamefont {Bl{\"u}gel}}, \bibinfo {author}
  {\bibfnamefont {F.}~\bibnamefont {Freimuth}}, \bibinfo {author}
  {\bibfnamefont {G.}~\bibnamefont {G{\'e}ranton}}, \bibinfo {author}
  {\bibfnamefont {M.}~\bibnamefont {Gibertini}}, \bibinfo {author}
  {\bibfnamefont {D.}~\bibnamefont {Gresch}}, \bibinfo {author} {\bibfnamefont
  {C.}~\bibnamefont {Johnson}}, \bibinfo {author} {\bibfnamefont
  {T.}~\bibnamefont {Koretsune}}, \bibinfo {author} {\bibfnamefont
  {J.}~\bibnamefont {{Iba{\~n}ez-Azpiroz}}}, \bibinfo {author} {\bibfnamefont
  {H.}~\bibnamefont {Lee}}, \bibinfo {author} {\bibfnamefont {J.-M.}\
  \bibnamefont {Lihm}}, \bibinfo {author} {\bibfnamefont {D.}~\bibnamefont
  {Marchand}}, \bibinfo {author} {\bibfnamefont {A.}~\bibnamefont {Marrazzo}},
  \bibinfo {author} {\bibfnamefont {Y.}~\bibnamefont {Mokrousov}}, \bibinfo
  {author} {\bibfnamefont {J.~I.}\ \bibnamefont {Mustafa}}, \bibinfo {author}
  {\bibfnamefont {Y.}~\bibnamefont {Nohara}}, \bibinfo {author} {\bibfnamefont
  {Y.}~\bibnamefont {Nomura}}, \bibinfo {author} {\bibfnamefont
  {L.}~\bibnamefont {Paulatto}}, \bibinfo {author} {\bibfnamefont
  {S.}~\bibnamefont {Ponc{\'e}}}, \bibinfo {author} {\bibfnamefont
  {T.}~\bibnamefont {Ponweiser}}, \bibinfo {author} {\bibfnamefont
  {J.}~\bibnamefont {Qiao}}, \bibinfo {author} {\bibfnamefont {F.}~\bibnamefont
  {Th{\"o}le}}, \bibinfo {author} {\bibfnamefont {S.~S.}\ \bibnamefont
  {Tsirkin}}, \bibinfo {author} {\bibfnamefont {M.}~\bibnamefont
  {Wierzbowska}}, \bibinfo {author} {\bibfnamefont {N.}~\bibnamefont
  {Marzari}}, \bibinfo {author} {\bibfnamefont {D.}~\bibnamefont {Vanderbilt}},
  \bibinfo {author} {\bibfnamefont {I.}~\bibnamefont {Souza}}, \bibinfo
  {author} {\bibfnamefont {A.~A.}\ \bibnamefont {Mostofi}},\ and\ \bibinfo
  {author} {\bibfnamefont {J.~R.}\ \bibnamefont {Yates}},\ }\bibfield  {title}
  {\bibinfo {title} {Wannier90 as a community code: New features and
  applications},\ }\href {https://doi.org/10.1088/1361-648X/ab51ff} {\bibfield
  {journal} {\bibinfo  {journal} {Journal of Physics: Condensed Matter}\
  }\textbf {\bibinfo {volume} {32}},\ \bibinfo {pages} {165902} (\bibinfo
  {year} {2020})}\BibitemShut {NoStop}%
\bibitem [{\citenamefont {Souza}\ \emph {et~al.}(2001)\citenamefont {Souza},
  \citenamefont {Marzari},\ and\ \citenamefont
  {Vanderbilt}}]{souzaMaximallyLocalizedWannier2001}%
  \BibitemOpen
  \bibfield  {author} {\bibinfo {author} {\bibfnamefont {I.}~\bibnamefont
  {Souza}}, \bibinfo {author} {\bibfnamefont {N.}~\bibnamefont {Marzari}},\
  and\ \bibinfo {author} {\bibfnamefont {D.}~\bibnamefont {Vanderbilt}},\
  }\bibfield  {title} {\bibinfo {title} {Maximally localized {{Wannier}}
  functions for entangled energy bands},\ }\href
  {https://doi.org/10.1103/PhysRevB.65.035109} {\bibfield  {journal} {\bibinfo
  {journal} {Physical Review B}\ }\textbf {\bibinfo {volume} {65}},\ \bibinfo
  {pages} {035109} (\bibinfo {year} {2001})}\BibitemShut {NoStop}%
\bibitem [{\citenamefont {{von Neuman}}\ and\ \citenamefont
  {Wigner}(1929)}]{vonneumanUberMerkwurdigeDiskrete1929}%
  \BibitemOpen
  \bibfield  {author} {\bibinfo {author} {\bibfnamefont {J.}~\bibnamefont {{von
  Neuman}}}\ and\ \bibinfo {author} {\bibfnamefont {E.}~\bibnamefont
  {Wigner}},\ }\bibfield  {title} {\bibinfo {title} {Uber merkw{\"u}rdige
  diskrete {{Eigenwerte}}. {{Uber}} das {{Verhalten}} von {{Eigenwerten}} bei
  adiabatischen {{Prozessen}}},\ }\href@noop {} {\bibfield  {journal} {\bibinfo
   {journal} {Physikalische Zeitschrift}\ }\textbf {\bibinfo {volume} {30}},\
  \bibinfo {pages} {467} (\bibinfo {year} {1929})}\BibitemShut {NoStop}%
\bibitem [{\citenamefont {Shtoff}\ and\ \citenamefont
  {R{\'e}rat}(2003)}]{shtoffFloquetGaugeinvariantCoupled2003}%
  \BibitemOpen
  \bibfield  {author} {\bibinfo {author} {\bibfnamefont {A.~V.}\ \bibnamefont
  {Shtoff}}\ and\ \bibinfo {author} {\bibfnamefont {M.}~\bibnamefont
  {R{\'e}rat}},\ }\bibfield  {title} {\bibinfo {title} {Floquet gauge-invariant
  coupled perturbation theory in calculations of the optical susceptibilities
  of molecules},\ }\href {https://doi.org/10.1134/1.1570473} {\bibfield
  {journal} {\bibinfo  {journal} {Optics and Spectroscopy}\ }\textbf {\bibinfo
  {volume} {94}},\ \bibinfo {pages} {496} (\bibinfo {year} {2003})}\BibitemShut
  {NoStop}%
\bibitem [{\citenamefont {{\v S}indelka}\ and\ \citenamefont
  {Moiseyev}(2007)}]{sindelkaFloquetPerturbationTheory2007}%
  \BibitemOpen
  \bibfield  {author} {\bibinfo {author} {\bibfnamefont {M.}~\bibnamefont {{\v
  S}indelka}}\ and\ \bibinfo {author} {\bibfnamefont {N.}~\bibnamefont
  {Moiseyev}},\ }\bibfield  {title} {\bibinfo {title} {Floquet perturbation
  theory: {{Applicability}} of the finite level approximation in different
  gauges},\ }\href {https://doi.org/10.1103/PhysRevA.76.043844} {\bibfield
  {journal} {\bibinfo  {journal} {Physical Review A}\ }\textbf {\bibinfo
  {volume} {76}},\ \bibinfo {pages} {043844} (\bibinfo {year}
  {2007})}\BibitemShut {NoStop}%
\bibitem [{\citenamefont
  {Suzuki}(1985)}]{suzukiDecompositionFormulasExponential1985}%
  \BibitemOpen
  \bibfield  {author} {\bibinfo {author} {\bibfnamefont {M.}~\bibnamefont
  {Suzuki}},\ }\bibfield  {title} {\bibinfo {title} {Decomposition formulas of
  exponential operators and {{Lie}} exponentials with some applications to
  quantum mechanics and statistical physics},\ }\href
  {https://doi.org/10.1063/1.526596} {\bibfield  {journal} {\bibinfo  {journal}
  {Journal of Mathematical Physics}\ }\textbf {\bibinfo {volume} {26}},\
  \bibinfo {pages} {601} (\bibinfo {year} {1985})}\BibitemShut {NoStop}%
\bibitem [{\citenamefont {Yates}\ \emph {et~al.}(2007)\citenamefont {Yates},
  \citenamefont {Wang}, \citenamefont {Vanderbilt},\ and\ \citenamefont
  {Souza}}]{yatesSpectralFermiSurface2007}%
  \BibitemOpen
  \bibfield  {author} {\bibinfo {author} {\bibfnamefont {J.~R.}\ \bibnamefont
  {Yates}}, \bibinfo {author} {\bibfnamefont {X.}~\bibnamefont {Wang}},
  \bibinfo {author} {\bibfnamefont {D.}~\bibnamefont {Vanderbilt}},\ and\
  \bibinfo {author} {\bibfnamefont {I.}~\bibnamefont {Souza}},\ }\bibfield
  {title} {\bibinfo {title} {Spectral and {{Fermi}} surface properties from
  {{Wannier}} interpolation},\ }\href
  {https://doi.org/10.1103/PhysRevB.75.195121} {\bibfield  {journal} {\bibinfo
  {journal} {Physical Review B}\ }\textbf {\bibinfo {volume} {75}},\ \bibinfo
  {pages} {195121} (\bibinfo {year} {2007})}\BibitemShut {NoStop}%
\bibitem [{\citenamefont {{Iba{\~n}ez-Azpiroz}}\ \emph
  {et~al.}(2018)\citenamefont {{Iba{\~n}ez-Azpiroz}}, \citenamefont {Tsirkin},\
  and\ \citenamefont {Souza}}]{ibanez-azpirozInitioCalculationShift2018}%
  \BibitemOpen
  \bibfield  {author} {\bibinfo {author} {\bibfnamefont {J.}~\bibnamefont
  {{Iba{\~n}ez-Azpiroz}}}, \bibinfo {author} {\bibfnamefont {S.~S.}\
  \bibnamefont {Tsirkin}},\ and\ \bibinfo {author} {\bibfnamefont
  {I.}~\bibnamefont {Souza}},\ }\bibfield  {title} {\bibinfo {title} {{\emph{Ab
  Initio}} calculation of the shift photocurrent by {{Wannier}}
  interpolation},\ }\href {https://doi.org/10.1103/PhysRevB.97.245143}
  {\bibfield  {journal} {\bibinfo  {journal} {Physical Review B}\ }\textbf
  {\bibinfo {volume} {97}},\ \bibinfo {pages} {245143} (\bibinfo {year}
  {2018})}\BibitemShut {NoStop}%
\bibitem [{\citenamefont
  {Higham}(2002)}]{highamAccuracyStabilityNumerical2002}%
  \BibitemOpen
  \bibfield  {author} {\bibinfo {author} {\bibfnamefont {N.~J.}\ \bibnamefont
  {Higham}},\ }\href {https://doi.org/10.1137/1.9780898718027} {\emph {\bibinfo
  {title} {Accuracy and {{Stability}} of {{Numerical Algorithms}}}}},\ \bibinfo
  {edition} {1st}\ ed.\ (\bibinfo  {publisher} {{Society for Industrial and
  Applied Mathematics}},\ \bibinfo {address} {Philadelphia},\ \bibinfo {year}
  {2002})\BibitemShut {NoStop}%
\end{thebibliography}%
\end{document}